\documentclass{article}
\pdfoutput=1 

\usepackage{jinstpub} 
\usepackage{caption}
\usepackage{subcaption}

\title{Effect of hole geometry on charge sharing and other parameters in GEM-based detectors}

\author[a,b]{Promita Roy,}
\author[c,1]{Purba Bhattacharya, \note{Corresponding author.}}
\author[a,b]{Prasant Kumar Rout,}
\author[a,b]{Supratik Mukhopadhyay}
\author[a,b]{and Nayana Majumdar}

\affiliation[a]{Saha Institute of Nuclear Physics,\\AF block, sector 1, Bidhannagar, Kolkata - 700064, India}
\affiliation[b]{Homi Bhabha National Institute,\\Anushaktinagar, Mumbai-400094, India}
\affiliation[c]{Department of Physics, School of Basic and Applied Sciences, Adamas University,\\Adamas Knowledge City, Barasat-Barrackpore Road, Kolkata, 700126, India}

\emailAdd{purba.bhattacharya1985@gmail.com}

\abstract{Gas Electron Multipliers (GEM) are among the more prominent Micro-Pattern Gaseous Detectors (MPGDs) and widely used in high energy particle physics experiments and various related applications.
Adoption of different production techniques lead to holes of varying geometries in GEM foils.
Since the response of a GEM-based detector is closely related to the hole geometry through the influence of the latter on charge sharing and transport through GEM foils, attempts have been made to relate hole configurations to different figures of merit of a detector.
Numerical simulations have been performed to study the effects of hole geometry on important parameters such as charge sharing, collection efficiency, extraction efficiency, gain, possibility of transition from avalanche to streamer modes for single, double and triple layer GEM detectors.
The numerical estimates have been compared to available experimental data.
The comparisons, although not always in agreement, are found to be generally encouraging.}


\arxivnumber{xxxx.xxxxx} 

\begin{document}
\maketitle
\flushbottom

\section{Introduction}
\label{sec:intro}

With the advent of lithography and printed circuit technology, there has been a vast development in the field of MPGDs \cite{Anton}.
Due to their good position and time resolution, high rate capability and radiation hardness, MPGDs are widely used in high energy particle physics experiments and related applications.
MPGDs relying on hole-based multiplication has been the subject of numerous studies because of the inherent property of dissociating volumes for electron amplification and signal collection.
GEM is one such MPGD introduced by Fabio Sauli in 1997 \cite{Sauli1997}.
GEM-based detectors, usually employing several layers of GEM-foils, are being widely used for both fundamental studies and societal applications due to their excellent performance and cost effectiveness.
For instance, GEM based readout chambers are being used for the upgrade of detectors such as CMS Forward Muon Spectrometer (FMS) \cite{CMSGEM}, ALICE Time Projection Chamber (TPC) \cite{ALICETPC}, medical imaging \cite{Sauli2016} and muon tomography \cite{Kondo}.
Quality Assurance (QA) of GEM foils is naturally very crucial in such experiments and has various aspects to be examined by different approaches \cite{Andrey, Jeremie, Brucken2019, Kalliskoski, Brucken2021}.

The key element of a GEM is a thin polyimide foil of a 50 $\mu$m thickness, coated with a 5 $\mu$m thick copper layer on both sides.
An array of bi-conical holes are chemically etched in a precise patterned structure with typical distances (pitch) of 140 $\mu$m.
However, manufacturing these symmetric bi-conical GEM holes requires complex double-mask technology \cite{Sauli1997}, where photo masks are applied separately on each side to create symmetric (70/50/70 $\mu$m) bi-conical holes.
This approach can lead to problems related to hole misalignment, especially for foils of large dimensions.
The single-mask technology \cite{Alfonsi, Gabriele, Marco, Serge} was proposed as an alternative where only one side is etched by applying a photo mask and the other side is electro-etched \cite{SergePatent}.
This technique leads to either conical \cite{Alfonsi}, or asymmetric (85/50/70 $\mu$m and 70/50/85 $\mu$m) \cite{Andrey}, but better aligned holes.
Being relatively easier and less expensive, the latter is steadily gaining in popularity and is being pursued by various experiments which need production of large GEM foils.

In short, it turns out that advantage of single-mask GEMs over double-mask GEMs is that the holes in the former are not misaligned, and disadvantage lies in the fact that single-mask bi-conical holes are usually not symmetric.
Thus, it has become important to understand the effect of these possible advantages and disadvantages on detector response before deciding on the technology to be preferred for a given experiment.
Among several recent investigations oriented towards this goal, effect of hole misalignment due to double-mask technique on the detector performance has been studied using optical scanning technique in a set of three GEM foils in \cite{Brucken2021}.
The Scanning White Light Interferometer (SWLI) was used in \cite{Brucken2021} as a basis to study the variation of GEM foil thickness for detailed 3D QA analysis of the GEM foil hole geometry.
Similarly, there has been a few study on single-mask hole geometry and their uniformity and how they affect the overall performance of the detector \cite{Sauli2016}.
Studies in \cite{Aashaq, Jeremie} and \cite{Keller} shows comparison of response of triple GEM detectors built using single- and double-mask GEM-foils.
The detector gain was found to be quite significantly affected by the hole geometry and also the orientation of the foil.
Such dependence was also observed for conical holes \cite{Alfonsi, Gabriele}.

Almost all the studies mentioned above are experimental in nature except \cite{Keller, Othmane} where good amount of numerical simulation is found.
The present work relies almost entirely on numerical models to explain the experimental observations reported previously, and to predict their expected behaviour in unexplored configurations.
As reported in the more recent studies \cite{Aashaq, Jeremie, Keller, Othmane}, along with the symmetric (SYM) bi-conical holes  (70/50/70 $\mu$m), two configurations of asymmetric bi-conical holes (diameters 85/50/70 and 70/50/85 $\mu$m) have been considered.
The first geometry represents double-mask GEMs, while the latter two geometries represent single mask GEMs.
The single-mask asymmetric configurations are termed as orientation A (OA) and orientation B (OB).
The major emphasis on these earlier studies had been on the variation of gain with the applied voltage.
Important differences were noticed in trends among the measurements, reasons for which are difficult to ascertain.
For example, while results from \cite{Keller} showed that detector gain for SYM GEMs is larger than both OA and OB, \cite{Jeremie} found that OB is larger than both SYM and OA.
According to \cite{Jeremie}, the average ratio of OB to OA gain values is quite large (3.6), while that from \cite{Aashaq} is smaller, close to 2.
Differences in spacing and other experimental parameters have been indicated by \cite{Keller} as the possible sources of the discrepancies but no specific study has been reported in this relation.
Another problem is the fact that the environmental parameters are found to vary from one set of data to another, making the measurements difficult to compare.

In this work, single-, double- and triple-layer GEM-based detectors built using the SYM, OA and OB hole geometries have been considered.
Parameters of single- and double-layer detectors have been chosen to match experimental observations reported in \cite{Bachmann}.
For triple-layer detectors, geometrical and electrical configurations matching \cite{Aashaq, Jeremie} (TG1), as well as \cite{Keller, Othmane} (TG2) have been studied.
Of particular interest has been the exploration of the way charge sharing among the GEM holes gets affected by these different configurations.
According to the present investigations, this parameter has far-reaching consequences influencing gain and transition from avalanche to streamer mode.

Occurrence of discharges is one of the major issues faced by MPGDs because these small gap detectors can be easily damaged by discharges \cite{PeskovFonte}.
The impact of discharges are, in general, negative for all detectors although the severity may vary.
For example, in GEM-based detectors, since the readout is not directly connected to the amplification device, discharges occurring in the latter may not cripple the detector itself.
Some of the other detectors, such as Micromegas, are not likely to be as resilient to discharges.
In any case, occurrence of frequent discharges is considered detrimental to reliable detector response of most MPGDs.

Initiation of streamer mode operation is usually considered as a precursor to the occurrence of discharges in gaseous detectors.
In the present work, effects of the hole geometry on the transition from avalanche to streamer mode have been studied in reasonable detail.
Since streamer mode operation involves localized accumulation of electrons / ions, charge sharing plays an important role in determining this transition via its effect on localization.
Thus, the role of charge sharing has been explored in this context.

It is conventional to invoke a limit akin to the Raether limit \cite{Raether} in order to ascertain streamer mode operation of a gaseous detector.
The limit is not a fixed number, but is influenced by the structure of the detector, among other factors, as mentioned in \cite{Fonte1998, Gasik, Prasant2021a}.
In particular, it was mentioned in \cite{Fonte1998} that higher gain could be achieved in multiple-layer GEM-based detectors due to enhanced diffusion of electrons.
Usually, the presence of electron number exceeding around $10^6 - 10^7$ is found to trigger the transition to streamer mode.

This approach has been slightly modified in the present work following the simulations carried out in \cite{Prasant2021a}.
In \cite{Prasant2021a}, it was observed that streamers occurring within a GEM hole are predominantly positive streamers in which the distortion of the applied electric field is due to the accumulation of ions, rather than electrons, within a GEM hole.
This observation is also supported by the knowledge that electrons, due to their much larger drift velocity and diffusion, are more likely to get quickly transported from one location to another.
The ions are relatively immobile, leading to the possibility of large concentration of ions in a small volume, especially where the ionizations take place.
In light of this experience, attempts have been made to identify the transition from avalanche to streamers in terms of the number of ions in a given hole.

Two independent approaches, particle and hydrodynamic (also referred to as the fluid model), have been pursued to carry out the present simulations.
Particle simulation of charge transport have been carried out using free and open source codes, while hydrodynamic simulation have been carried out using a commercially available Finite Element Method (FEM) package.
The hydrodynamic simulation is, in fact, a hybrid model since it uses several inputs obtained from tools using particle approach.
Various pros and cons of these two fundamentally different modeling approaches have been explored and discussed.

Brief details of the numerical models used in this work have been discussed in section \ref{NumModel}.
Since the electric field is expected to play an important role in determining the detector response, electric fields in different geometrical configurations have been presented in section \ref{EField}.
Section \ref{ChargeSharing} describes the impact of different experimental parameters on the sharing of charge among different GEM holes.
Collection and extraction efficiency of these detectors are also described in this section.
The variation of gain, avalanche and streamer mode operations are investigated in section \ref{ResDis} along with relevant discussions relating charge sharing and space charge to these final observable.
Finally, the concluding remarks are presented in section \ref{ConRem} .

\section{Numerical model}
\label{NumModel}

\subsection{Geometry and gaseous mixture}
\label{Geometry}
GEM foils are assumed to be made of 50 $\mu$m thick Kapton sandwiched between copper layers of 5 $\mu$m thickness.
The pitch between the holes is considered to be 140 $\mu$m.

\begin{figure}[htbp]
	\centering 
	\includegraphics[width=.4\textwidth]{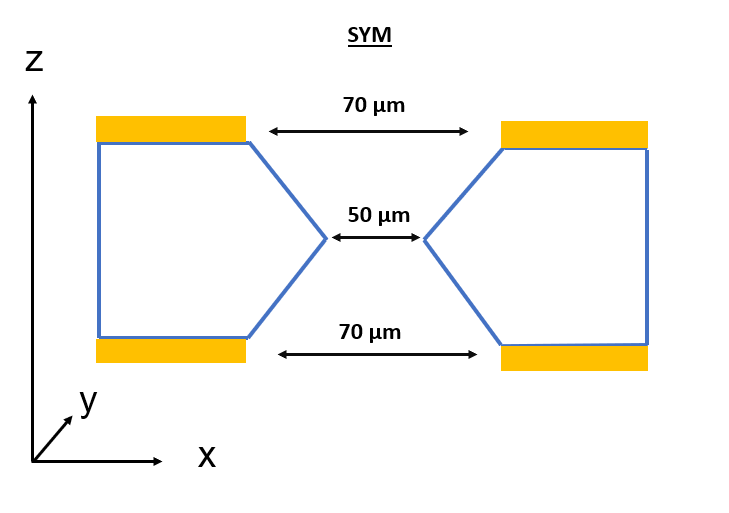}
	\qquad
	\caption{\label{SYM} Schematic hole geometry for a double-mask GEM foil (SYM) and the coordinate system.}
\end{figure}

\begin{figure}[htbp]
	\centering

	\begin{subfigure}{0.5\textwidth}
	\centering
	\includegraphics[width=.8\linewidth]{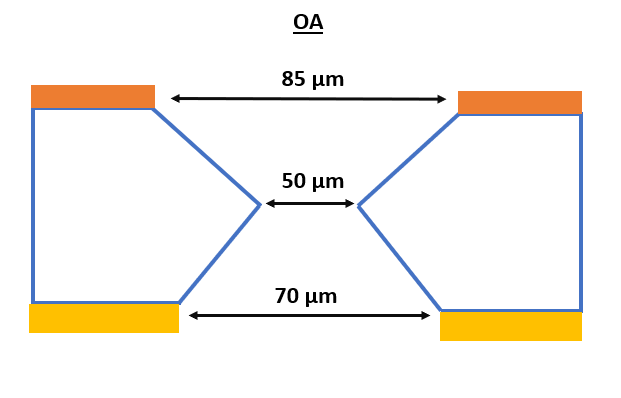}
	\caption{}
	\end{subfigure}%
	\begin{subfigure}{0.5\textwidth}
	\centering
	\includegraphics[width=.8\linewidth]{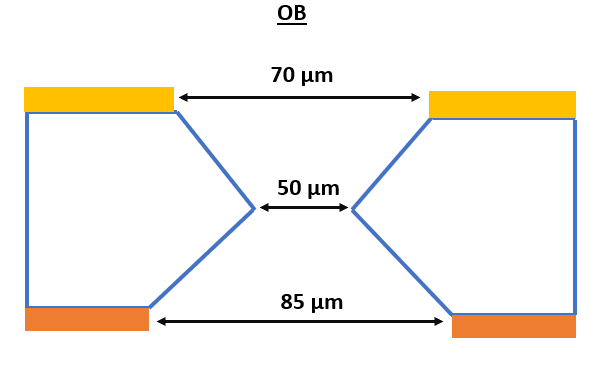}
	\caption{}
	\end{subfigure}

	\caption{Hole orientations (a) OA and (b) OB of single-mask GEM foils.}
	\label{OAvsOB} 
\end{figure}

Double-mask bi-conical symmetric hole (SYM) is as shown in figure \ref{SYM}.
OA has a larger opening towards the drift cathode whereas OB has a larger opening towards the readout as illustrated in figure\ref{OAvsOB}.

It may be noted here that, assuming 140 $\mu$m pitch for all the cases, optical transparency of a 70 $\mu$m hole is around 20\%, while that of a 85 $\mu$m hole is considerably larger at around 30\%.

Single GEM detectors having  3mm drift and 1mm induction gaps has been simulated. 
Double GEM detectors are assumed to have one 1mm induction gap added to the single GEM geometry.
For these detectors, G1 is the GEM foil that is adjacent to the drift volume and G2 is the one adjacent to the induction volume.
For triple GEM detectors, two different geometries have been considered in order to compare with \cite{Aashaq} and \cite{Keller}.
The former has drift:transfer1:transfer2:induction as 2mm:2mm:2mm:2mm in order to be able to invert the detector and obtain OB from OA, and vice versa, easily.
The latter has 3mm:1mm:2mm:1mm (drift:transfer1:transfer2:induction) geometry in agreement with the CMS configuration \cite{CMSGEM}.
For triple GEMs, G1 will mean the gem foil adjacent to the drift volume, G2 will refer to the middle GEM foil which has transfer gaps on either side and the GEM foil adjacent to the induction volume will be referred to as G3.

\begin{figure}[htbp]
	\centering 
	\includegraphics[width=.5\textwidth]{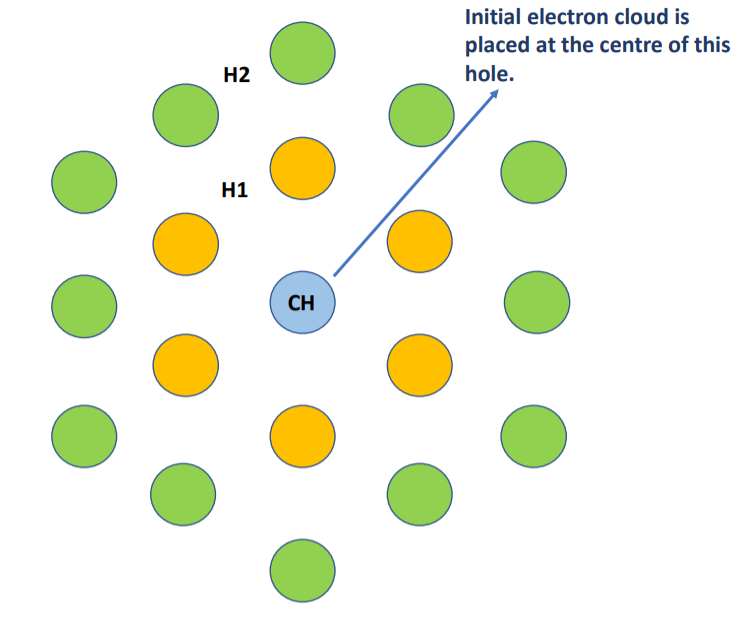}
	\caption{\label{CentralHoleHexagons} Schematic representation of central hole and hexagonal holes of a GEM foil.}
\end{figure}

In order to study the charge sharing among different GEM holes, the geometry shown in figure \ref{CentralHoleHexagons} has been considered.
For the purpose of discussion, the central hole will be referred to as CH, first hexagon as H1 and so on.

A gaseous mixture of argon(Ar)-carbon dioxide(CO$_2$) in volume proportion 70:30 is assumed to be present in the detector for all the simulations presented in this work.

\subsection{Electrical configuration}
\label{EConfig}

Results from several existing works \cite{Bachmann, Jeremie, Aashaq, Keller, Othmane} have been used to establish the numerical estimates obtained during the present investigation.
The electrical configurations of the numerical detectors created for the current simulations have been matched accordingly.

\subsection{Mathematical Models}
\label{MathModel}
The free, open source Garfield \cite{Garfield} framework is used here for simulating detector response using the particle approach.
It incorporates HEED \cite{Heed} to estimate the amount of primary ionization and MAGBOLTZ \cite{Magboltz} for estimating electron transport parameters including Townsend and attachment coefficients.
Electric field has been computed using neBEM \cite{neBEM1, neBEM2}, which once again is free, open source, and the commercially available FEM package COMSOL \cite{COMSOL}.
While the particle model  is realistic and extremely capable \cite{Purba2015, Purba2017}, it has certain limitations that turn out to be important especially when the number of charged particles is large.
Since the model inherently tracks each of the charged particles, the computation time becomes inordinately long in such cases.
This is even more true when attempts are made to incorporate space charge effects, which the particle model does not consider by default.

The hybrid model, on the other hand, uses HEED and MAGBOLTZ information to simulate dynamics of electron and ion species within a background neutral fluid medium utilizing the Transport of Dilute Species (TDS) interface of COMSOL.
The hydrodynamic model \cite{Prasant2021a, Prasant2021b}, by its very nature, does not attempt to follow charged particles as isolated and independent (although interacting) entities.
It considers the detector medium as a continuum (solvent), which a reasonable approximation, especially under normal conditions.
In addition, it considers the primary and secondary electrons and ions also as charged fluids (solutes) that are dilute in comparison to the background.
There are, naturally, several approximations involved that are reasonable only under certain special circumstances.
For example, the assumption of continuum for the charged species is acceptable when the density of electrons and ions are large which fails especially during the early phase when only the primaries are present.
However, it is straightforward to include the effects of space charge in this hydrodynamic model.
As a result the hybrid approach, in contrast with the particle approach, works well when the number of charged particles is large.

\section{Electric field}
\label{EField}

Electric field configuration within a detector volume, in addition to the gaseous mixture, plays a major role in determining the detector response.
Since the gas mixture was predetermined, an attempt was made to understand the effect of hole geometry leading to different field configurations.
It may be noted that two completely different approaches \cite{neBEM1, COMSOL} were used in this work in order to compute the electric field and they produced very comparable results.
Here, the effects of hole geometry on electric field variation for single, double and triple GEM detectors will be explored using neBEM as the field solver.
Drift and induction fields are as mentioned in \ref{EConfig}.

\begin{figure}[htbp]
\centering 

\begin{subfigure}{0.5\textwidth}
\centering
\includegraphics[width=.8\linewidth]{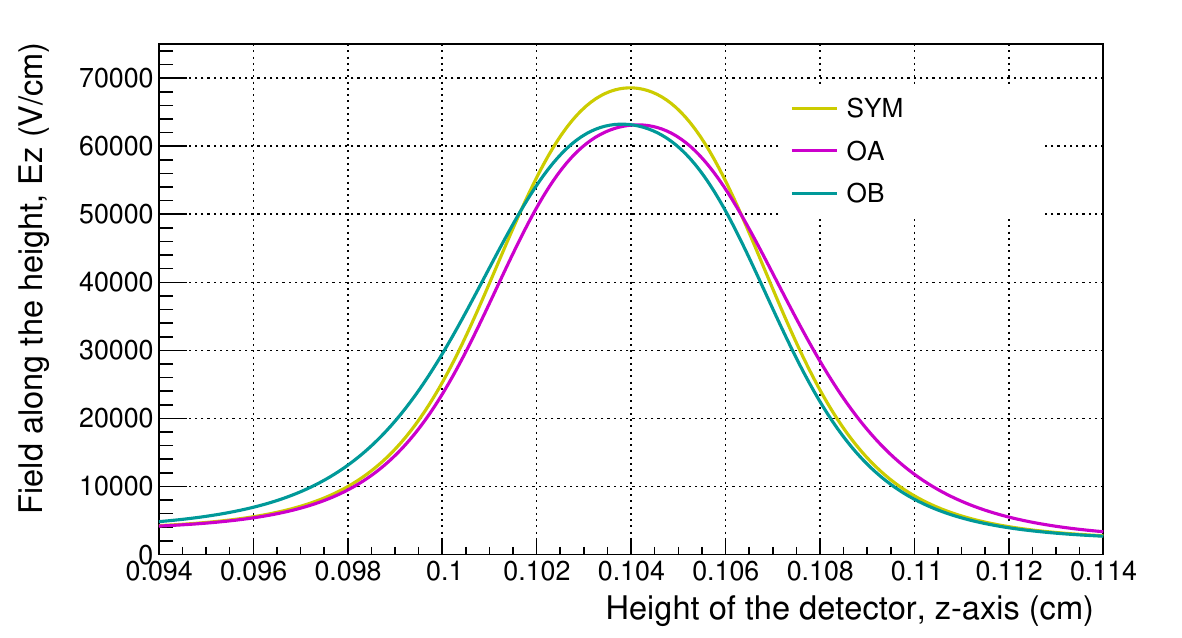}
\caption{}
\end{subfigure}%
\begin{subfigure}{0.5\textwidth}
\centering
\includegraphics[width=.8\linewidth]{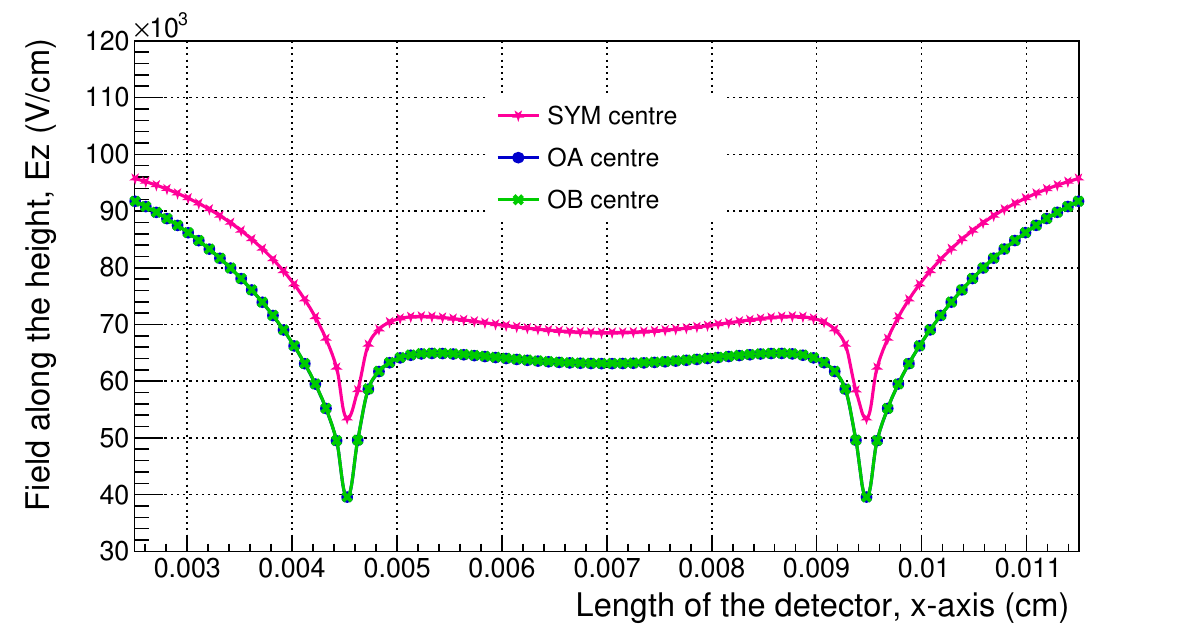}
\caption{}
\end{subfigure}

\begin{subfigure}{0.5\textwidth}
\centering
\includegraphics[width=.8\linewidth]{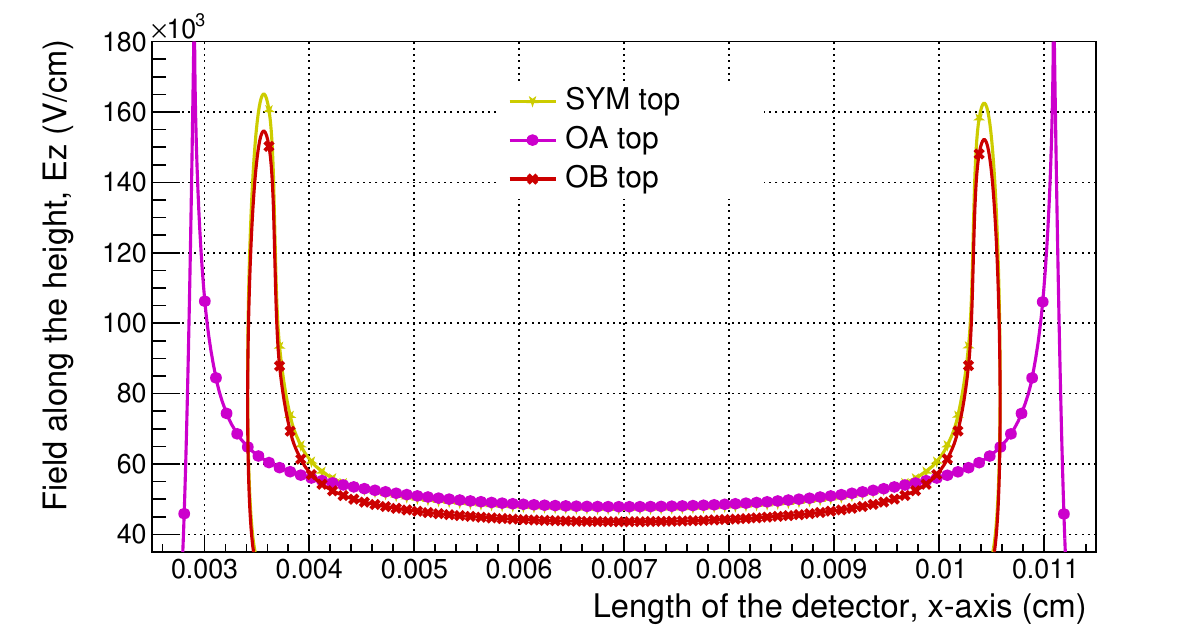}
\caption{}
\end{subfigure}%
\begin{subfigure}{0.5\textwidth}
\centering
\includegraphics[width=.8\linewidth]{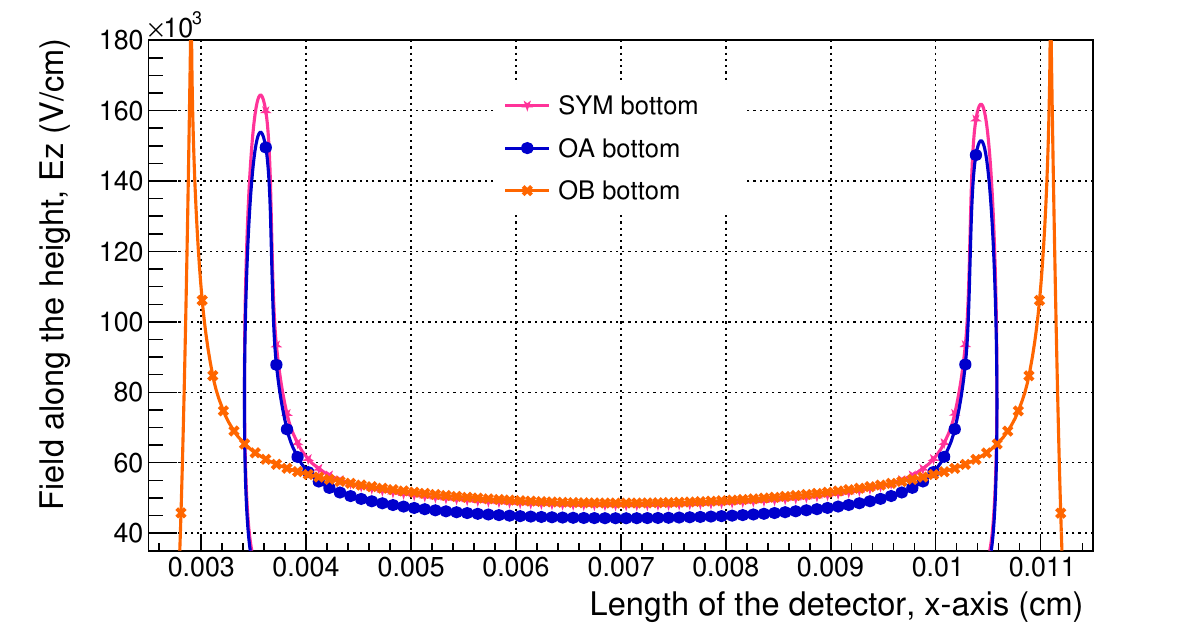}
\caption{}
\end{subfigure}

\caption{\label{EFieldSingleGEM} Electric field (E$_z$) for SYM, OA and OB configuration for a single GEM detector: (a) along Z, (b) along X at hole center, (c) along X at hole top and (d) along X at hole bottom.}
\end{figure}

From figure \ref{EFieldSingleGEM}(a), it is seen that electric field corresponding to a given GEM voltage along z-axis for OA and OB are equal and less than that of SYM.
The fields for OA and OB attain maximum value at position shifted towards the drift gap and towards the transfer gap, respectively.
The magnitude of the maximum field values for OA and OB are different but not very significantly so.
Total electric field for SYM is larger at the hole centre, in comparison to OA and OB, as shown in figure \ref{EFieldSingleGEM}(b).
It is very interesting to note that field along z-axis at the top of the holes is same for OA and SYM whereas electric field for OB is less than other two configuration (figure \ref{EFieldSingleGEM}(c)).
An exactly reverse situation is observed at the bottom of the holes where electric field for OA is less than that of SYM and OB as shown in figure \ref{EFieldSingleGEM}(d).

\begin{figure}[htbp]
\centering 

\begin{subfigure}{0.5\textwidth}
\centering
\includegraphics[width=.9\linewidth]{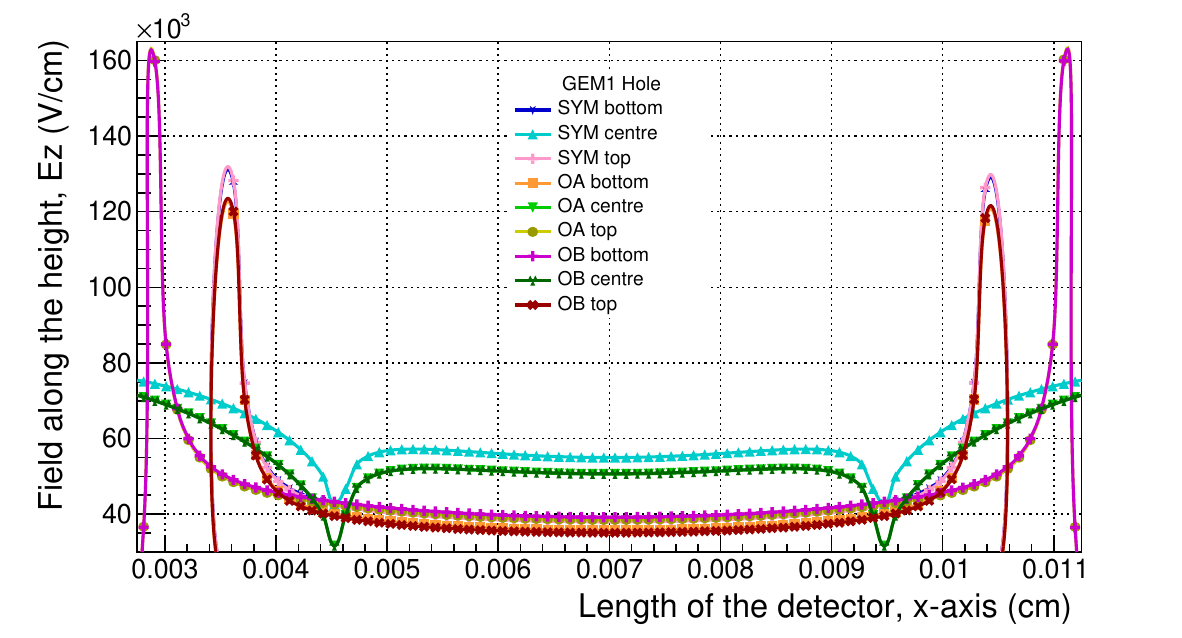}
\caption{}
\end{subfigure}%
\begin{subfigure}{0.5\textwidth}
\centering
\includegraphics[width=.9\linewidth]{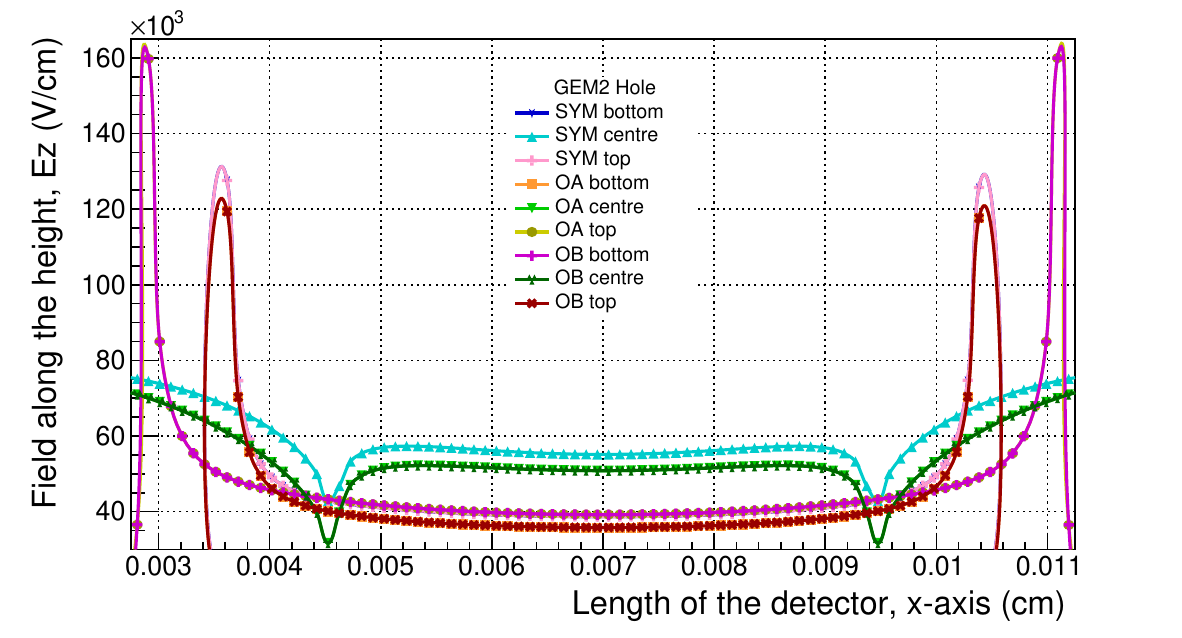}
\caption{}
\end{subfigure}

\begin{subfigure}{0.8\textwidth}
\centering
\includegraphics[width=.8\linewidth]{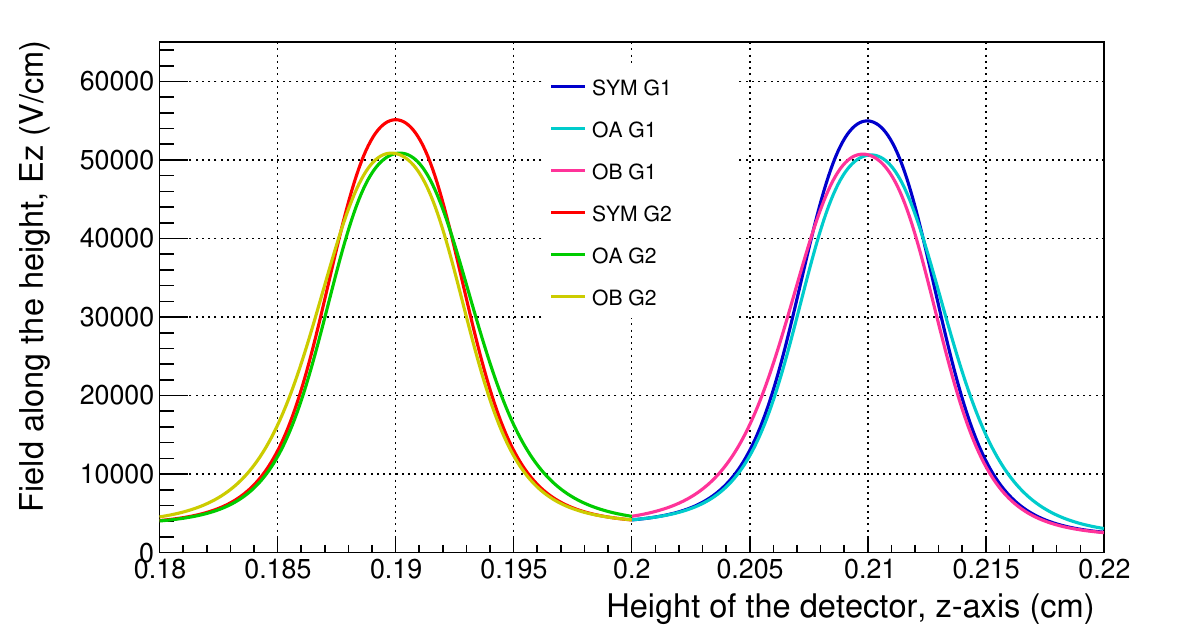}
\caption{}
\end{subfigure}

\caption{\label{EFieldDoubleGEM} Electric field (E$_z$) for SYM, OA and OB configurations for a double GEM detector: (a) along x for G1, (b) along x for G2 and (c) along z for G1 and G2}
\end{figure}

An exercise similar to the above is next carried out for a double GEM configuration.
The reason for this repetition is the fact that the double GEM configuration has slightly different field value in the transfer zone, in comparison to the drift field of a single GEM.
It is interesting to see the effect of this difference on the eventual field values in and around the GEM holes.
In figures \ref{EFieldDoubleGEM} (a) and (b), variation of total electric field along x-axis have been shown.
These two figures are very similar since there is little difference in the electric configuration for G1 and G2 in this double GEM setup.
It can be observed that for both the GEMs G1 and G2, the electric field at the hole centre is larger for SYM and smaller (and almost identical) for OA and OB.
Next important pattern to observe is the fact that all the configurations have more or less the same total electric field at the top and bottom of the hole.

Figure \ref{EFieldDoubleGEM}(c) shows the variation of total electric field along the hole axis, z-axis for G1 and G2.
It should be noted here that, in order to facilitate comparison of the variation between G1 and G2, a z shift has been applied to the G2 so that the curves remain close to each other.
The magnitude of electric field for the SYM configuration is clearly larger than OA and OB.
Moreover, it may once again be noted that due to the geometrical asymmetry, the fields for OA and OB do not attain maximum value at the hole centre.
Rather, the peak is offset by a small but significant amount for OA towards the drift gap and OB towards the transfer gap for G1.
Similarly, for G2, the maximum is attained towards the transfer gap for OA and towards the induction gap for OB.
The magnitude of the maximum field values for OA and OB are different but not very significantly so.
Due to this subtle tilt of the field profile along the hole axis, the OA configuration has higher field values towards the drift gap for G1 and towards the transfer gap for G2.
The opposite is true for the OB configuration.

It may be mentioned here that at these field values, the drift velocity, longitudinal and transverse diffusion coefficients do not show significant variations, for the gas mixture under consideration.
As a result, the mentioned variations are unlikely to affect charge sharing, collection and extraction efficiencies.
The same is not true for the Townsend ($\alpha$) and attachment ($\eta$) coefficients, however.
For both single-layer and multiple-layer GEM detectors, SYM has the highest field values.
As a result, the effective Townsend coefficient ($\alpha - \eta$) will be maintained at a higher level for SYM, in comparison to both OA and OB.
OA will have more $\alpha - \eta$ towards drift / transfer gap, and less $\alpha - \eta$ towards transfer / induction gap, while OB will have less $\alpha - \eta$ towards drift / transfer gap, but more $\alpha - \eta$ towards transfer / induction gap.
Since more ionization occurs towards the extraction zone, the OB configuration seems to be more favored than OA, as far as $\alpha - \eta$ is considered.

\section{Charge sharing related studies}
\label{ChargeSharing}

\subsection{Charge sharing}
Since the pitch between GEM holes are of the same order as electron diffusion in the drift and transfer volumes, the sharing of electrons among GEM holes is perceptible.
Thus, charge sharing is expected to play a significant role in determining response of hole-based gaseous detectors in general, and GEM-based detectors, in particular. 
In this section, effects of hole geometry on charge sharing has been studied with a view to relate hole geometry to various figures of merit of the detector, including transition from avalanche to streamer mode operation.

In order to carry out this study, a single electron located on the CH axis has been released from 1mm above G1.

\begin{figure}[htbp]
\centering 

\begin{subfigure}{0.5\textwidth}
\centering
\includegraphics[width=.8\linewidth]{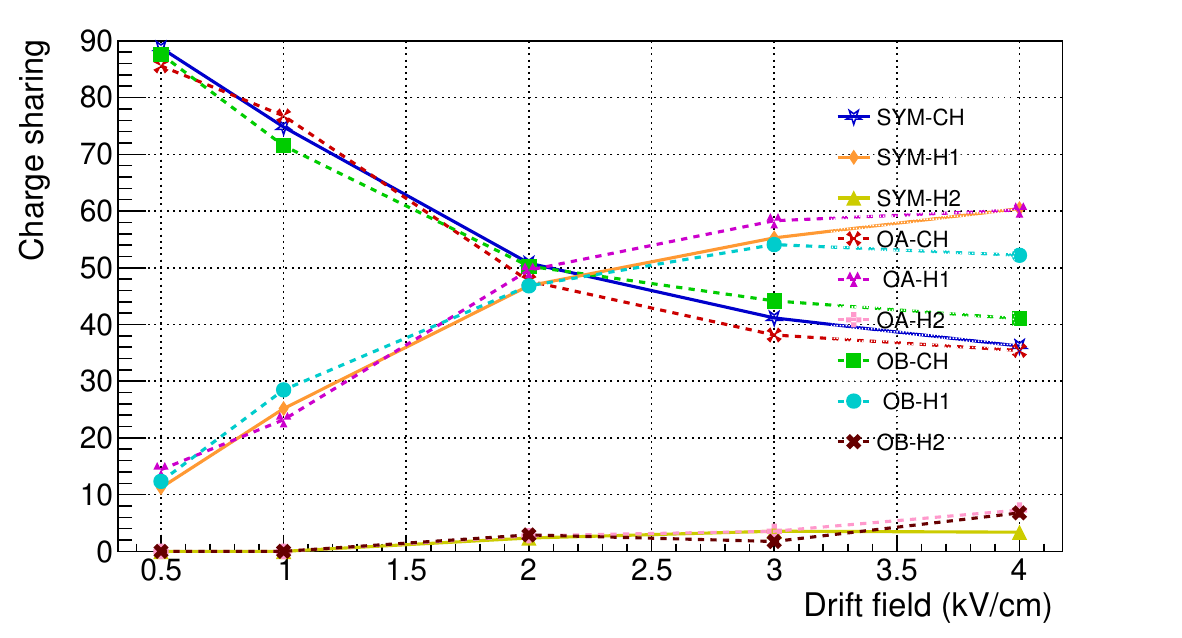}
\caption{}
\end{subfigure}%
\begin{subfigure}{0.5\textwidth}
\centering
\includegraphics[width=.8\linewidth]{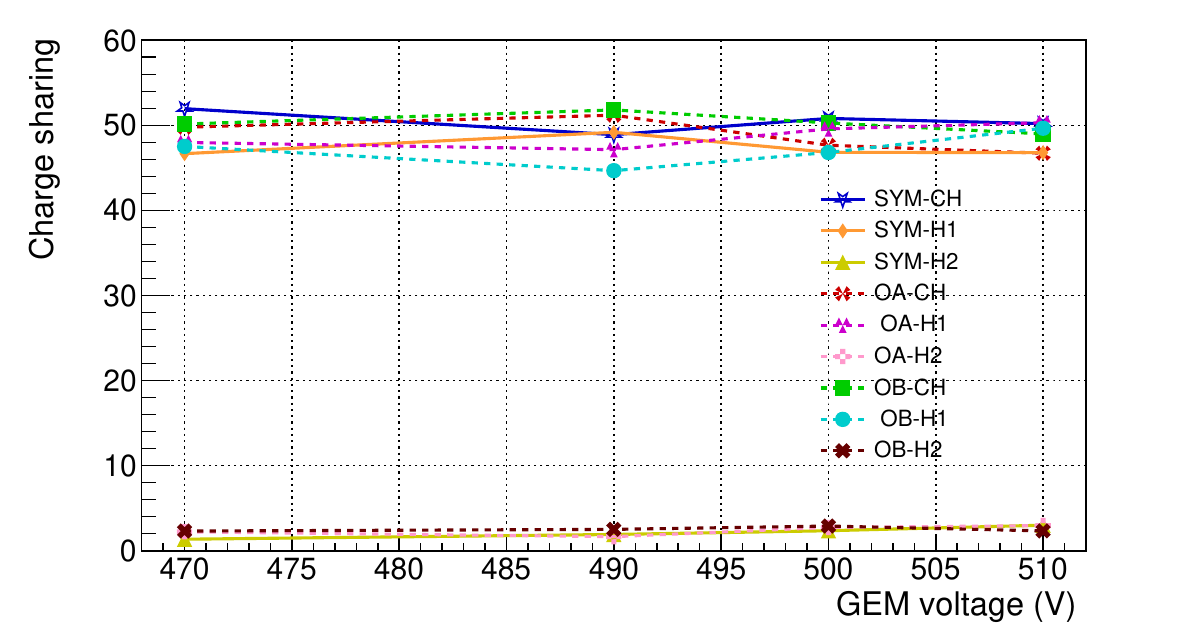}
\caption{}
\end{subfigure}

\caption{\label{ChrgShr1GEM} Sharing of charges (\%) for single GEM (a) drift field and (b) GEM voltage (SYM, OA, OB)}
\end{figure}

From figure \ref{ChrgShr1GEM}(a), it is observed that at lower drift field strengths (0.5kV/cm and 1kV/cm), around 70-85\% of total charge is the share of CH, and the remaining 15-30\% charge is received by H1.
As the drift field strength is increased, the share of CH decreases and that of H1 and H2 increase.
This observation is explained by the fact that the transverse diffusion coefficient increases considerably till the electric field is around 3kV/cm.
As the drift field approaches the value of 4kV/cm, the charge sharing values tend to saturate due to the same reason.
Beyond this value, the CH share for all the configurations are found to be around 40\%, while that of H1 is within 50-60\% and H2 less than 10\%.
This observation may be explained by the fact that both transverse and longitudinal diffusion coefficients reach respective plateau regions once electric field exceeds 3kV/cm.
Such a variation is almost absent as the voltage across the GEM foil is altered, as observed from figure \ref{ChrgShr1GEM}(b).

\begin{figure}[htbp]
\centering 

\begin{subfigure}{0.5\textwidth}
\centering
\includegraphics[width=.8\linewidth]{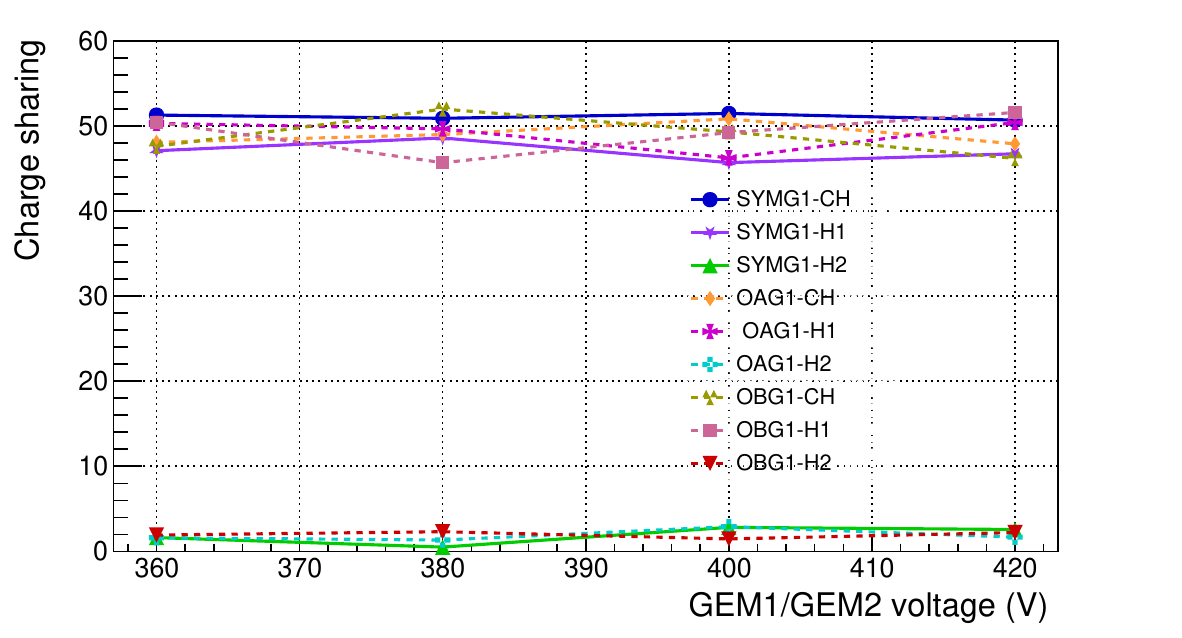}
\caption{}
\end{subfigure}%
\begin{subfigure}{0.5\textwidth}
\centering
\includegraphics[width=.8\linewidth]{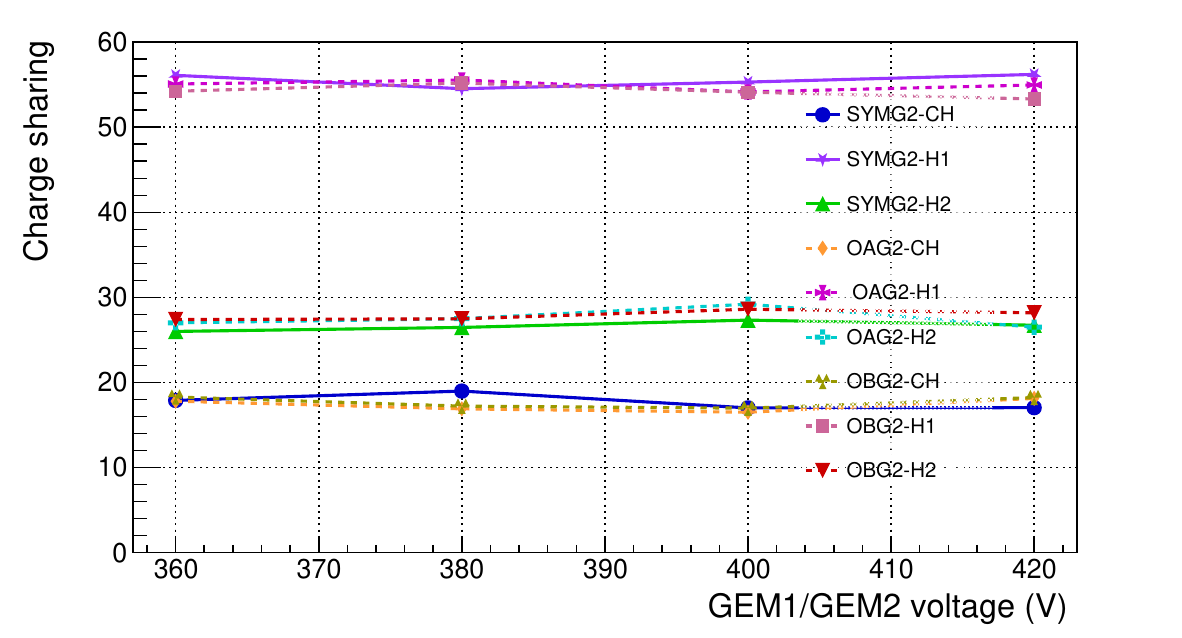}
\caption{}
\end{subfigure}

\caption{\label{ChrgShr2GEM} Sharing of charges (\%) for double GEM with different (a) G1 and (b) G2 voltages for all the three orientations (SYM, OA, OB)}
\end{figure}

After a second GEM layer is introduced, G1 retains the single-GEM values while the sharing increases further on G2, as shown in figure \ref{ChrgShr2GEM}.
As a result, for G1, while CH retains the same range of shared charge (50\%), H1 has a share below 45-50\% while the share for H2 increases above 5\%.
For G2, share of CH is less than 20\%, that of H1 is 50-55\% and that of H2 is close to 30\%.

\begin{figure}[htbp]
\centering 

\begin{subfigure}{0.5\textwidth}
\centering
\includegraphics[width=.8\linewidth]{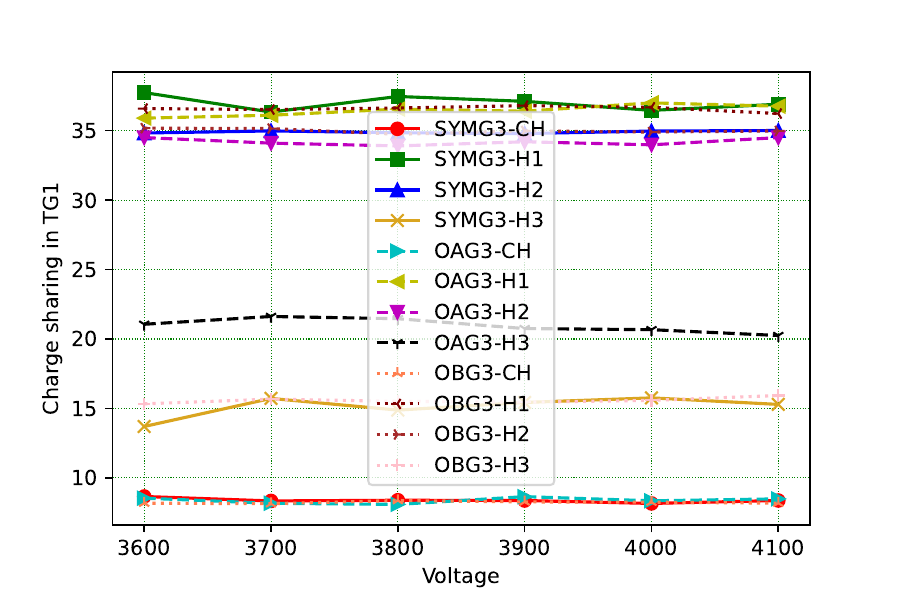}
\caption{}
\end{subfigure}%
\begin{subfigure}{0.5\textwidth}
\centering
\includegraphics[width=.8\linewidth]{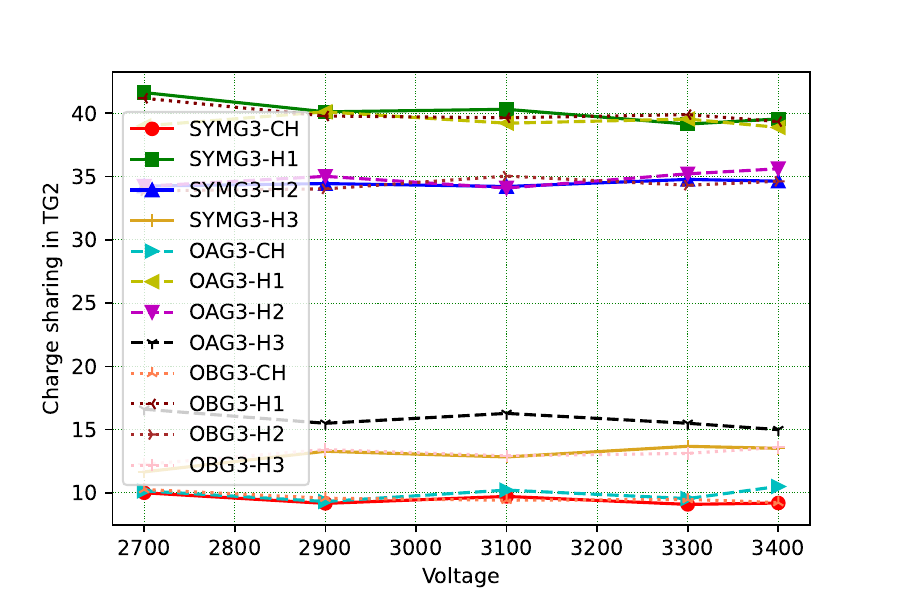}
\caption{}
\end{subfigure}

\caption{\label{ChrgShr3GEM} Sharing of charges (\%) on G3 for (a) TG1 and (b) TG2 with different drift voltages (in Volts) for all the three orientations (SYM, OA, OB)}
\end{figure}

Figure \ref{ChrgShr3GEM} shows the condition for triple-GEM configurations TG1 and TG2 on G3.
Here it is observed that, for TG1, CH has a share of around 8\%, both H1 and H2 has around 35\% share while the share for H3 varies from 15 to 20\%.
The OA configuration is found to have larger share of electrons in H3 (20\%).
For TG2, SYM has close to 10\% share in CH, 40\% in H1, 35\% in H2 and around 15\% in H3.
Once again, H3 has a larger share (around 18\%) for the OA configuration.
The fact that maximum share of electrons is almost always maintained by the central hole (around 10\% in comparision to 8\% in H1), plays a crucial role in our attempt to model transition from avalanche to discharge.
This allows the use of an axisymmetric model that reduces the burden of a full 3D simulation of this complex phenomenon.

\subsection{Collection efficiency}

Collection efficiency for a GEM foil adjacent to the drift volume is defined as
\begin{equation}
\eta_C = \frac{electrons~entering~GEM~hole}{primary~~electrons}
\label{CollEffEqn}
\end{equation}

\begin{figure}[htbp]
\centering 

\begin{subfigure}{0.5\textwidth}
\centering
\includegraphics[width=.8\linewidth]{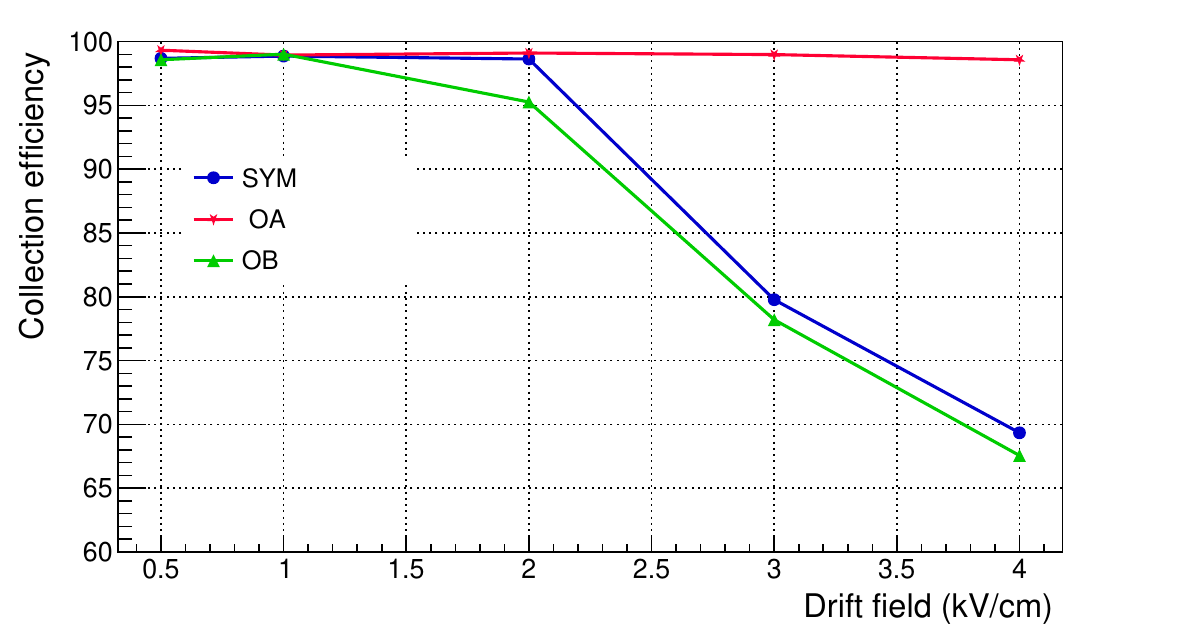}
\caption{}
\end{subfigure}%
\begin{subfigure}{0.5\textwidth}
\centering
\includegraphics[width=.8\linewidth]{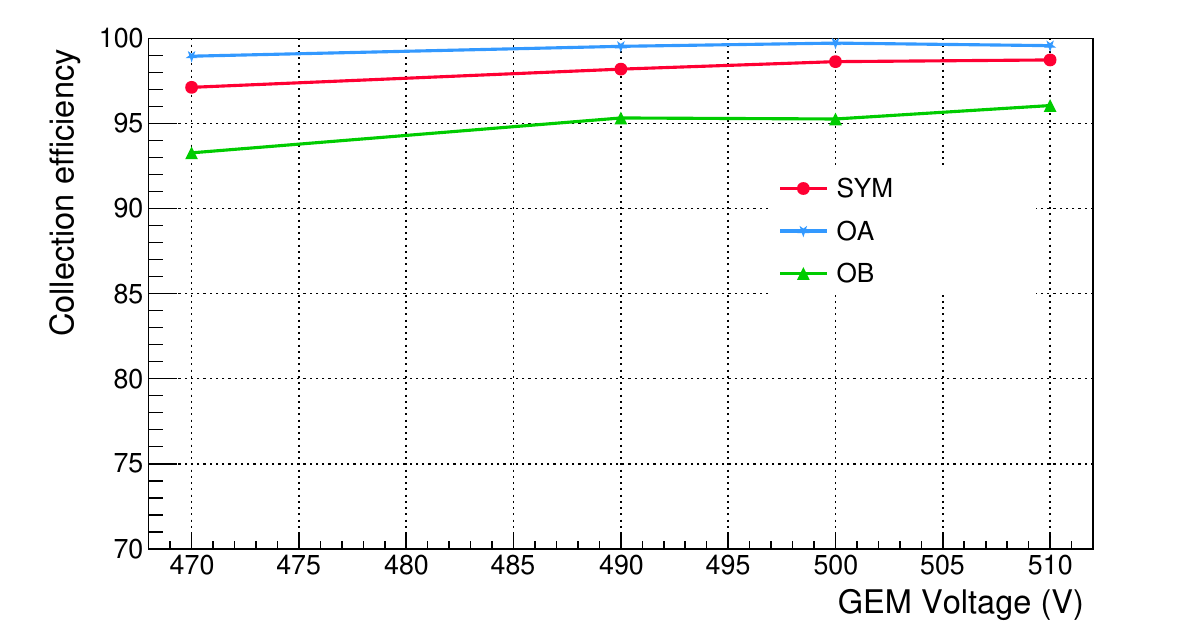}
\caption{}
\end{subfigure}

\caption{\label{CollEff1GEM} Variation of collection efficiency (\%) with (a) drift field and (b) GEM voltage for single GEM}
\end{figure}

Collection efficiency for two different single-mask orientations have been calculated and it is observed that OA with a bigger hole opening towards the drift cathode has almost 99$\%$ collection efficiency for single GEM (figure \ref{CollEff1GEM}), whereas collection efficiency for SYM and OB decreases with drift field as expected from earlier studies of double-mask GEMs.
This is clearly an advantage of OA which does not need a fine tuning of the drift and transfer fields, as far as collection of electrons is considered.
 
\begin{figure}[htbp]
\centering 

\begin{subfigure}{0.5\textwidth}
\centering
\includegraphics[width=.8\linewidth]{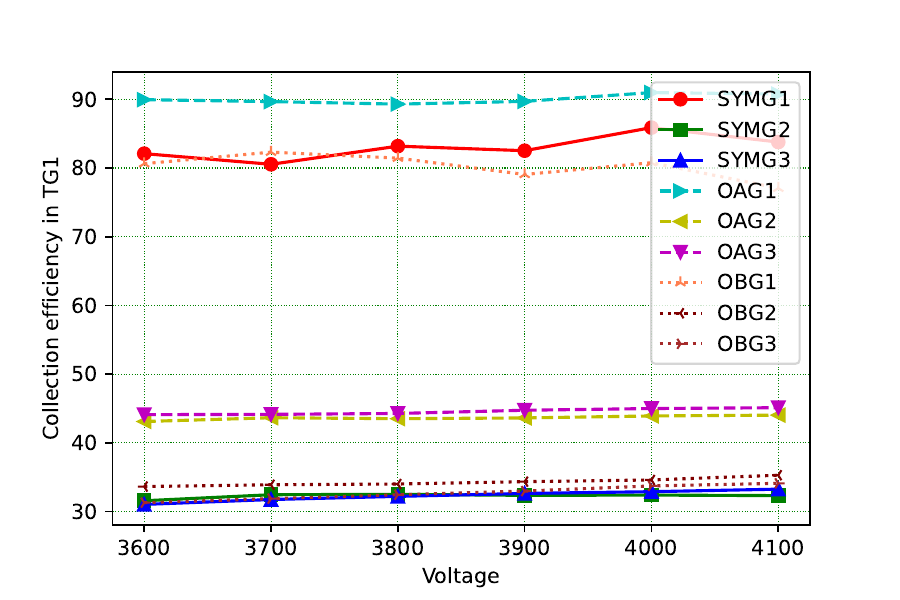}
\caption{}
\end{subfigure}%
\begin{subfigure}{0.5\textwidth}
\centering
\includegraphics[width=.8\linewidth]{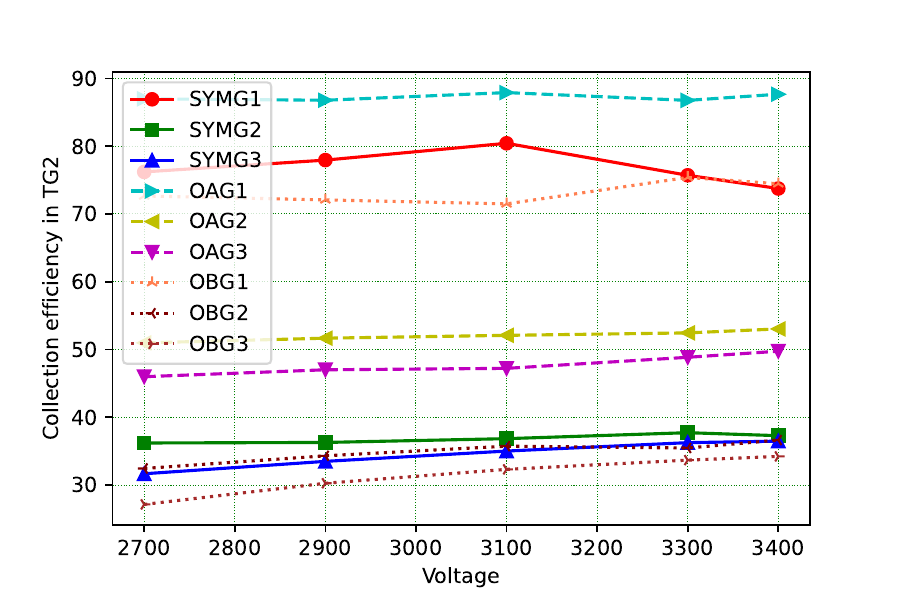}
\caption{}
\end{subfigure}

\caption{\label{CollEff3GEM} Variation of collection efficiency (\%) of G1, G2 and G3 for different drift voltages (in Volts) in (a) TG1 and (b) TG2}
\end{figure}

The same pattern is repeated for double and triple GEM detectors, as shown in figure \ref{CollEff3GEM}, for triple GEM ones.
The collection efficiency for OA is consistently higher than SYM and OB for all the GEM foils.

\subsection{Extraction efficiency}
Extraction efficiency for a GEM foil facing either a transfer, or induction gap is defined as
\begin{equation}
\eta_E = \frac{electrons~in~transfer~/~induction~volume}{total~electrons~in~GEM~hole}
\label{ExtEffEqn}
\end{equation}
The rest of the electrons are considered lost on the GEM anode.

\begin{figure}[htbp]
\centering 
\includegraphics[width=.6\textwidth]{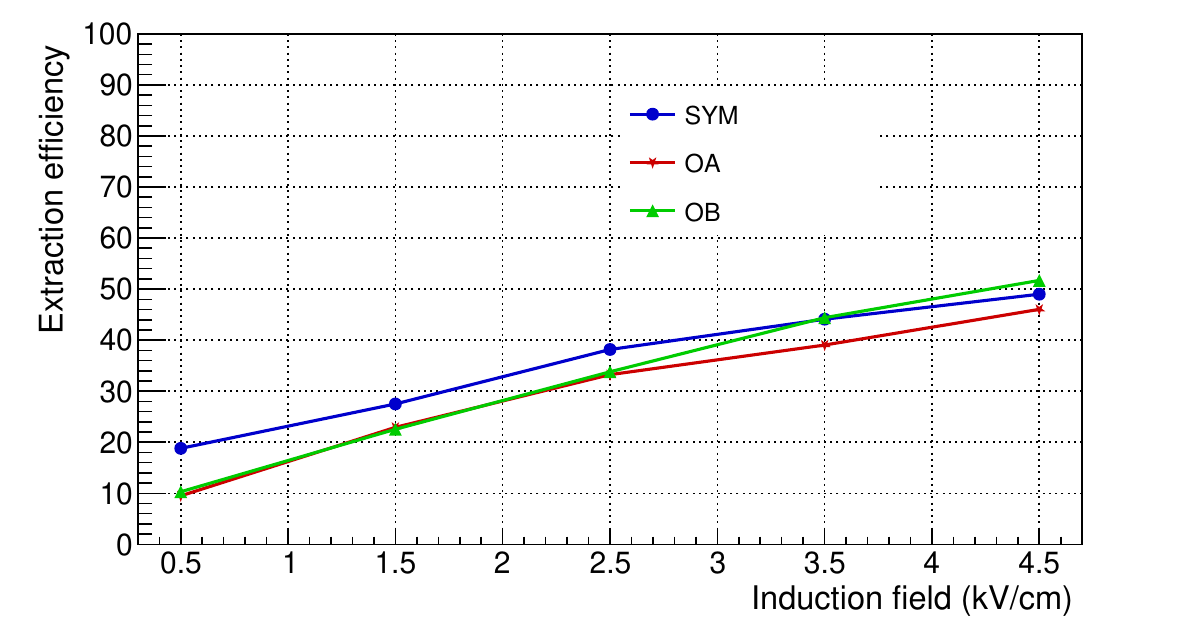}
\caption{\label{ExtEff1GEM} Variation of extraction efficiency (\%) with applied induction field for single GEM}
\end{figure}

Extraction efficiency has been calculated varying the induction field, as shown in figure \ref{ExtEff1GEM}.
The induction field has a strong effect on extraction efficiency, but no significant advantage of one geometry over the other is observed.
SYM has higher efficiency for lower values of the induction field, while for higher values of induction field, the OB configuration performs the best.
OA remains at a lower efficiency value for the entire range of induction field.

\begin{figure}[htbp]
\centering 

\begin{subfigure}{0.5\textwidth}
\centering
\includegraphics[width=.8\linewidth]{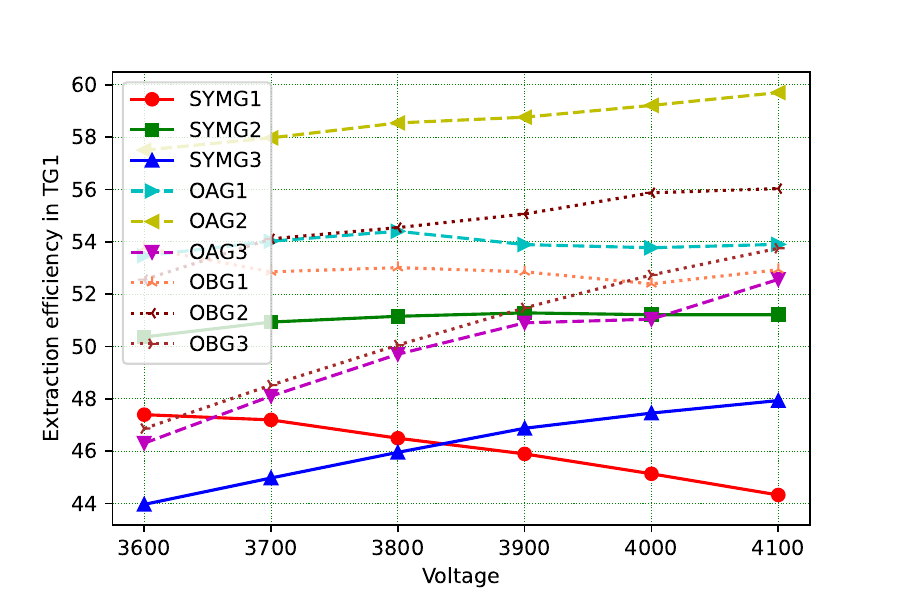}
\caption{}
\end{subfigure}%
\begin{subfigure}{0.5\textwidth}
\centering
\includegraphics[width=.8\linewidth]{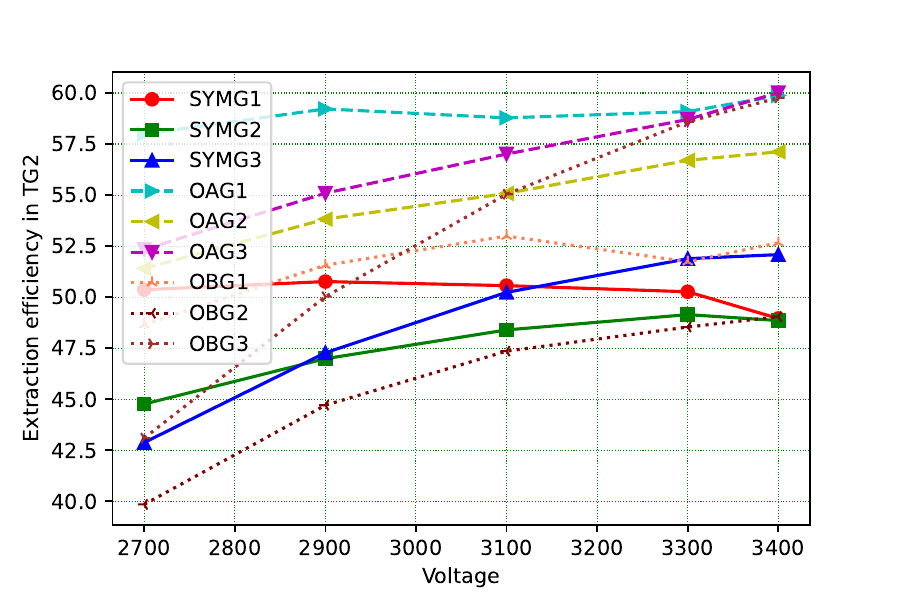}
\caption{}
\end{subfigure}

\caption{\label{ExtEff3GEM} Variation of extraction efficiency (\%) with drift voltage (in Volts) for all the GEM foils of (a) TG1 and (b) TG2}
\end{figure}

Variation of extraction efficiency for the triple GEM configurations are shown in figures \ref{ExtEff3GEM}(a) and (b).
It is observed that the SYM configuration has lower extraction efficiency among the three.
OA and OB has similar extraction efficiencies, with OA having having an edge over OB for both TG1 and TG2.
Once again it is observed that, within the range considered, the variation of voltage across the GEM foils do not affect the extraction efficiency to a significant amount, and the value for all the configurations remain in between 40 to 60\%.

\subsection{Transmission efficiency}
\begin{figure}[htbp]
\centering 

\begin{subfigure}{0.5\textwidth}
\centering
\includegraphics[width=.8\linewidth]{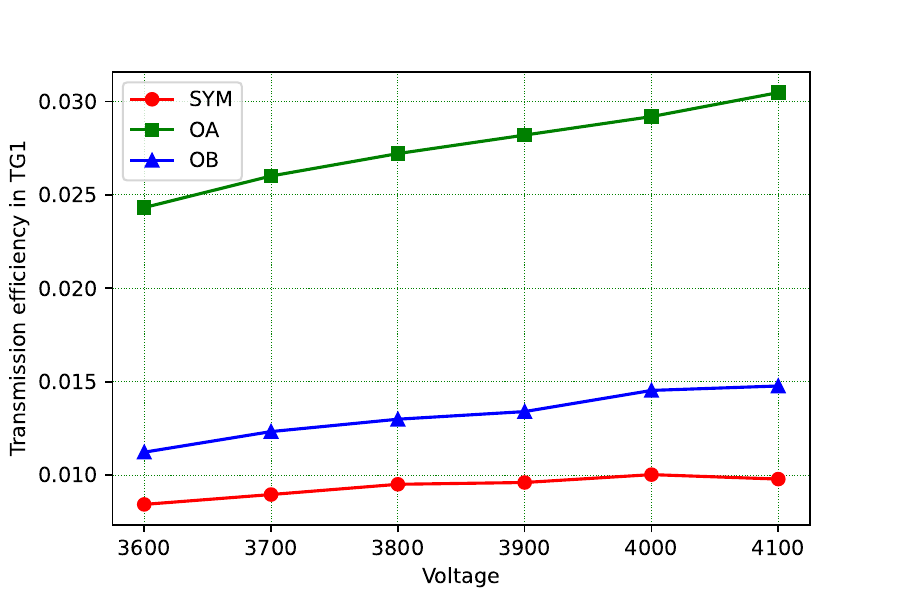}
\caption{}
\end{subfigure}%
\begin{subfigure}{0.5\textwidth}
\centering
\includegraphics[width=.8\linewidth]{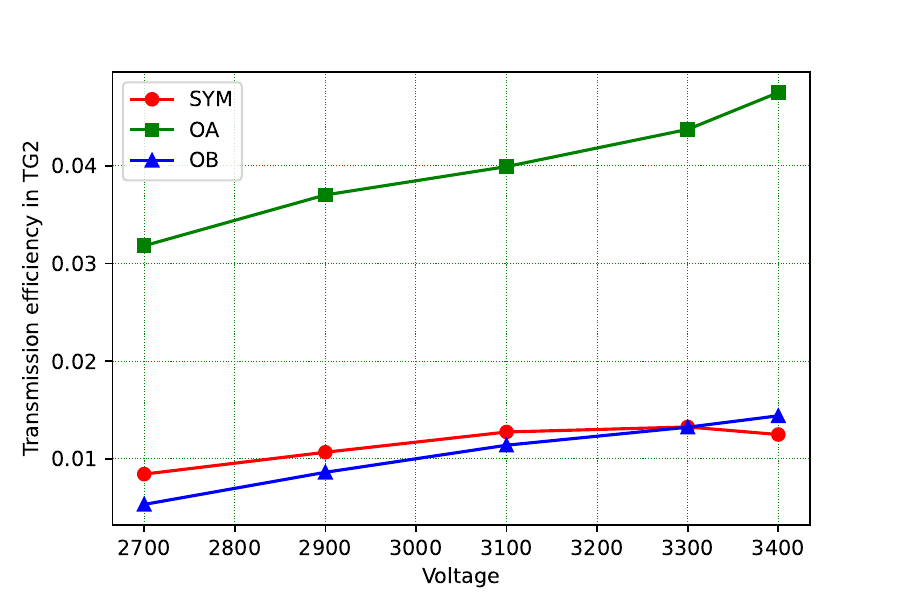}
\caption{}
\end{subfigure}

\caption{\label{TGTEff} Variation of transmission efficiency (\%) with drift voltage (in Volts) for (a) TG1 and (b) TG2}
\end{figure}

Effective transmission efficiency of TG1 and TG2 detector configurations have been estimated by multiplying relevant collection and extraction efficiencies.
Variation of transmission efficiency with applied drift voltage for these two detector configurations are shown in figure \ref{TGTEff}.
It can be observed that the OA configuration has significantly higher transmission efficiencies than SYM and OB for both TG1 and TG2.
For TG1, OB has larger efficiency of transmission than SYM, while for TG2, the efficiencies for both SYM and OB are simlar.

\section{Results and discussions}
\label{ResDis}
\subsection{Gain variation}
\begin{figure}[htbp]
\centering 
\includegraphics[width=.7\textwidth]{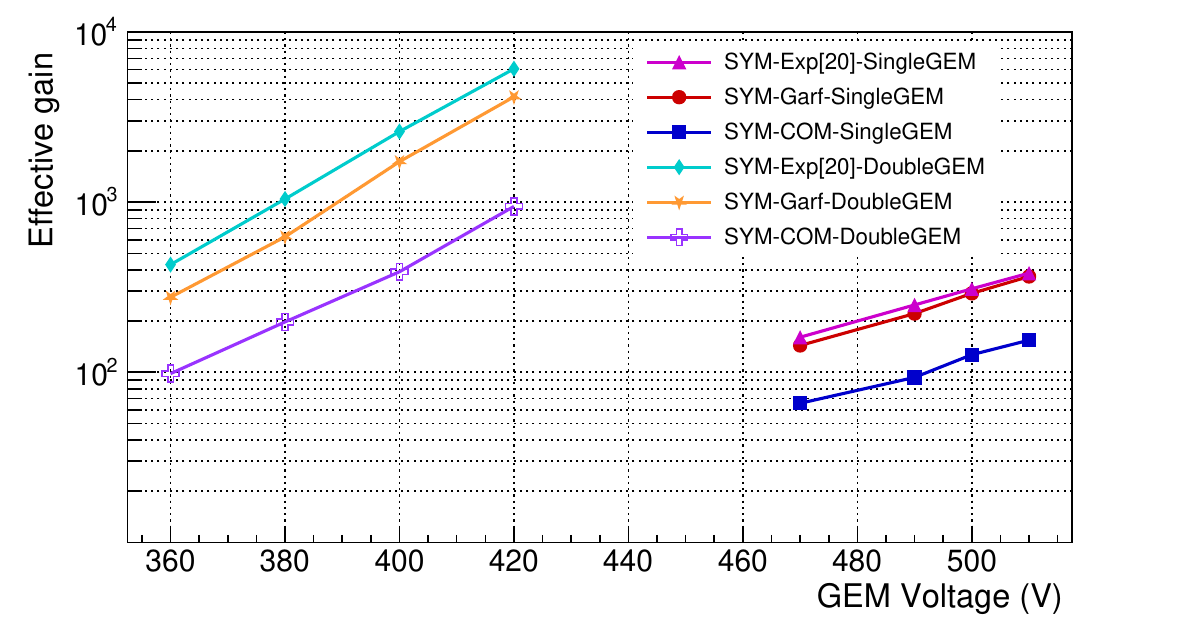}
\caption{\label{SingleDoubleGainExpSimSYM} Comparison of simulation and experimental results of gain for a single and double GEM.}
\end{figure}

Numerically estimated gain values using both hydrodynamic and particle models have been compared with experimental values \cite{Bachmann}, as shown in figure \ref{SingleDoubleGainExpSimSYM}.
It is observed that both the numerical approaches agree well with the experimental measurements, the fluid model estimates being lower than the particle model estimates.

\begin{figure}[htbp]
\centering 

\begin{subfigure}{0.5\textwidth}
	\centering
    \includegraphics[width=.8\linewidth]{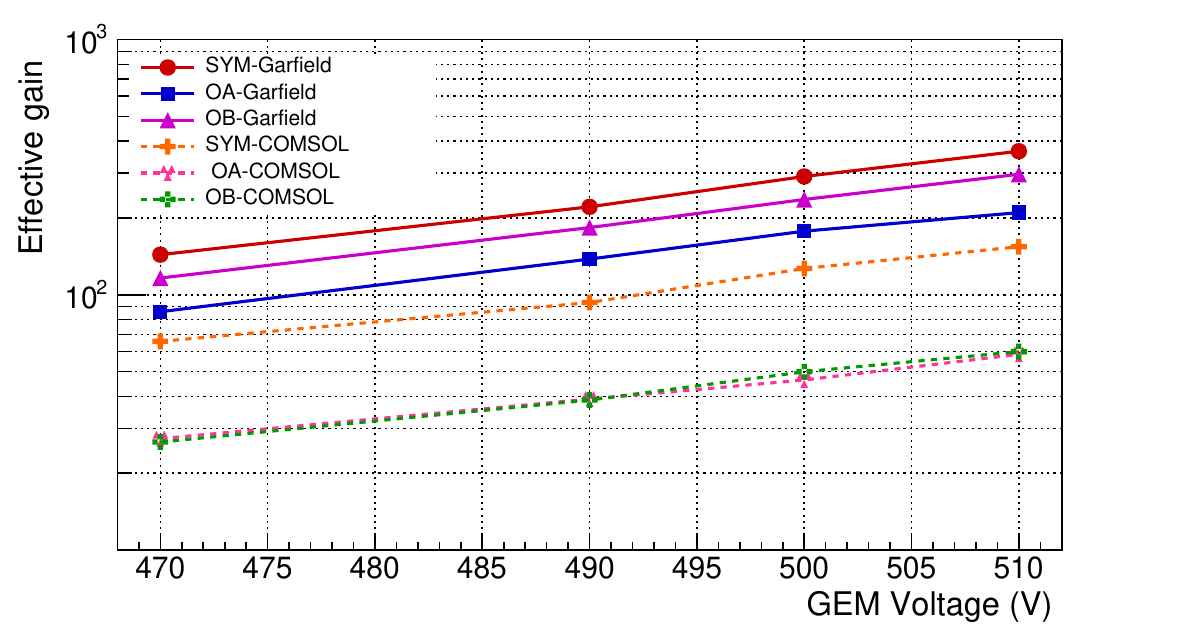}
    \caption{}
\end{subfigure}%
\begin{subfigure}{0.5\textwidth}
	\centering
	\includegraphics[width=.8\linewidth]{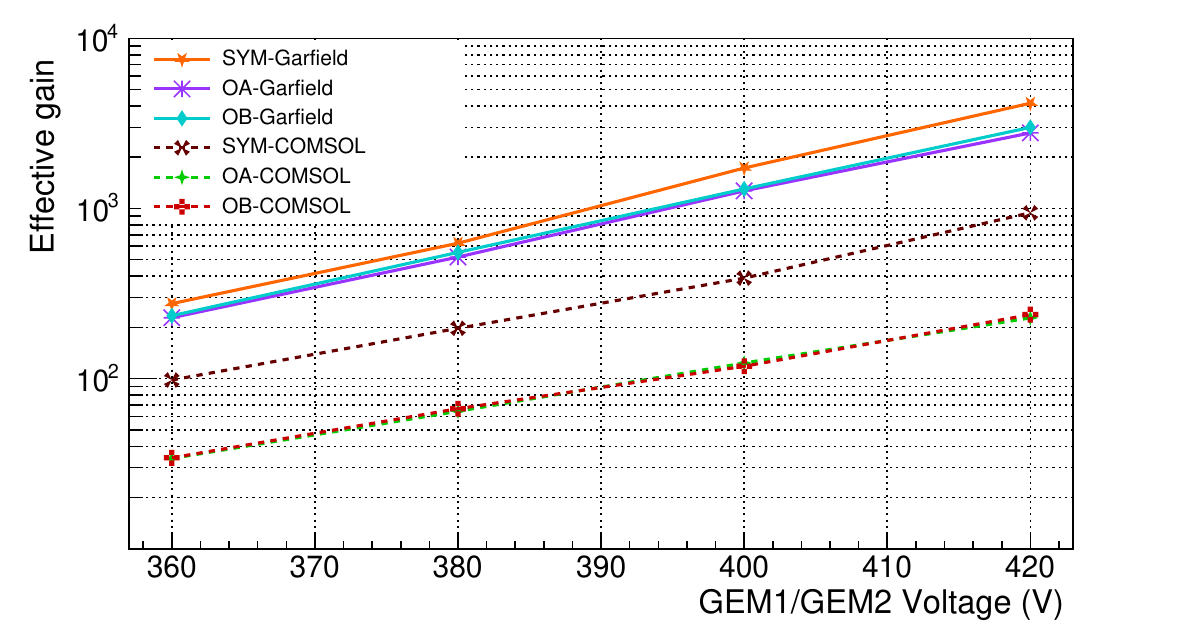}
	\caption{}
\end{subfigure}

\caption{\label{SingleDoubleGainSimSYMOAOB} Simulated gains for SYM, OA and OB configurations in (a) single and (b) double GEM detector.}
\end{figure}

In figure \ref{SingleDoubleGainSimSYMOAOB}, numerical estimations of single and double GEM detectors for SYM, OA and OB configurations have been presented.
For both the cases, SYM gain values are larger than both OA and OB gain value.
While using the particle approach, it is interesting to observe that gain for OB is larger that of OA configuration for a single GEM detector.
The gain of OA is slightly larger than that estimated for OB for a double GEM detector when using the same approach.
The difference in gain between OA and OB almost is non-existent when using the fluid approach.
The compounded effect of field distribution, collection and extraction efficiencies is likely to be the cause of these differences.

\begin{figure}[htbp]
\centering 

\begin{subfigure}{0.5\textwidth}
\centering
\includegraphics[width=.8\linewidth]{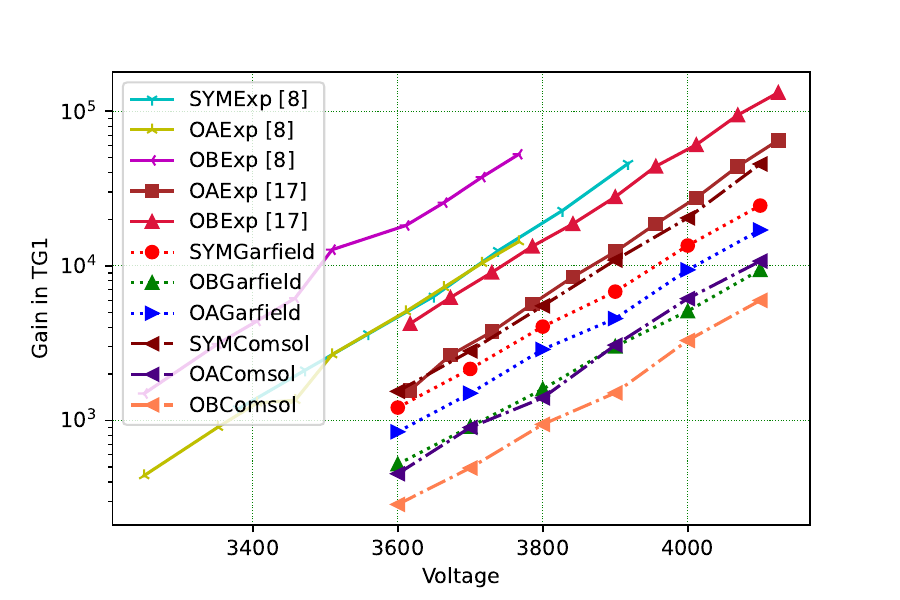}
\caption{}
\end{subfigure}%
\begin{subfigure}{0.5\textwidth}
\centering
\includegraphics[width=.8\linewidth]{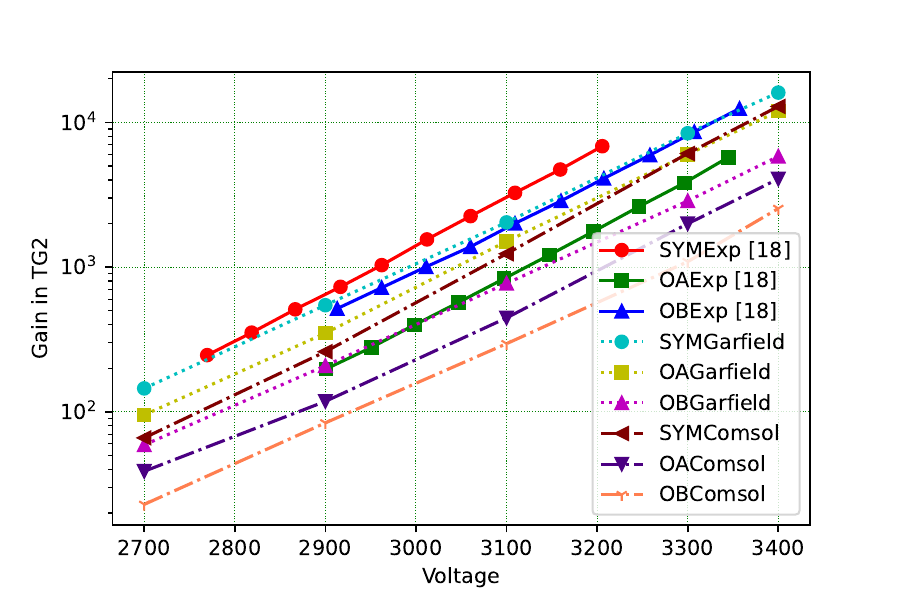}
\caption{}
\end{subfigure}

\caption{\label{TripleGEMGains} Comparison of measured and simulated gains versus drift voltage (in Volts) for (a) TG1 and (b) TG2.}
\end{figure}

Figures \ref{TripleGEMGains}(a) and (b) show comparisons among experimentally measured gain for triple GEM detectors with the numerical estimates obtained using both Garfield and COMSOL.
Both TG1 and TG2 configurations considered in these plots.
It is observed that experimentally TG1 finds OB to have the largest value of gain (Figure \ref{TripleGEMGains}(a)).
It needs to be noted here that results presented in \cite{Jeremie} and \cite{Aashaq} are qualitatively similar for OA and OB, but the voltage ranges and gain values are different.
They represent results obtained from two different experiments.
However, in order to reduce the number of plots while maintaining the essence of the observations, results for the SYM configuration from \cite{Jeremie} has been overlayed on the same plot.
It may be mentioned here that \cite{Aashaq} does not exhibit SYM measurements.
SYM and OA in \cite{Jeremie} produce very similar gain values, while OB is significantly higher.
OB is higher than OA for \cite{Aashaq} as well.
However, the ratio between OB and OA has been stated to be around 3.6 in \cite{Jeremie} and 2 by \cite{Aashaq}.
The simulations, on the contrary, indicate that gain for SYM should be the largest, followed by OA and OB.
This is true for both hydrodynamic and particle simulations, the latter being larger and more comparable to the experimental values.
In the case of TG2 \cite{Othmane}, it is observed experimentally that SYM has the largest gain, followed by OB and OA, which is supported by numerical simulations provided in the same reference.
The numerical simulations in the present work, on the contrary, indicate SYM to be the largest, followed by OA and OB as estimated by both particle and fluid approaches.
This apparent conflict between the results can be easily resolved by considering the influence of environmental pressure on the measurements and numerical estimates.

The numerical calculations presented in the present paper have been carried out at 985 mbar for all three configurations.
However, experimental and numerical simulation results presented in figure 8 of \cite{Othmane} show that SYM, OA and OB configurations have been operated at pressures 997 mbar, 986 mbar and 977 mbar respectively.
Moreover, figure 9(b) of \cite{Othmane} shows that a difference of 10 mbar pressure introduces significant change in effective detector gain.
Normalizing the numerical results presented in \cite{Othmane} to one particular pressure (for instance 986 mbar, close to one of the pressures used in this paper) indicate SYM to have the highest gain.
Gains from OA and OB configurations are found to be almost overlapping each other.
Also, normalizing the gain at one particular pressure increases the effective gain margin between SYM and OA/OB configurations.
However, the simulated gain values in \cite{Othmane} are almost 7 times smaller than the experimental measurements.
In contrast, the particle model numerical estimates presented here are around half of the experimentally measured values, while the fluid model estimates are close to one-fifth of them.

Finally, it may be concluded from the gain plots that for a given environmental condition, the SYM configuration has the largest gain for all the configurations.
Use of OA and OB configurations lead to similar gains, one being slightly more than the other, depending on the detector geometry.
This is despite the fact that the transmission efficiency of OA is more than SYM and OB.
SYM has larger field than both OA and OB throughout the GEM hole and OB has larger field than OA in the region of the GEM hole that is close to the transfer / induction gap.
This leads to more ionization for both SYM and OB than that in OA.
For SYM, it more than offsets the lack of transmission efficiency, while for OB it just cancels the transmission efficiency advantage of OA.
As a result, for a given voltage configuration, both OA and OB yields less gain, in comparison to SYM.
Gains for OA and OB are comparable, though.
This is not necessarily bad, as we will see in the next discussion.

\subsection{Transition from avalanche to streamer mode}
Next, attempts were made, following the approach described in \cite{Prasant2021a}, to ascertain the voltages near which transition from  avalanche to streamer occurs for each of the configurations.
It may be mentioned here that due to the presence of large number of charged particles and the likely importance of space charge, it is very difficult to pursue the particle model in this part of the investigation.
In \cite{Prasant2021a}, it was found that number of ions at which single GEM detector was evolving into streamer was around 9 $\times 10 ^6$.
The same number was close to 1.5 $\times 10^7$ and 2 $\times 10^7$ for double- and triple-GEM configurations.
This can be related to the fact that the number of charges inside the CH of single GEM is around 2-3 times that of double GEM and almost 5 times as large as that of a triple GEM.
As a result, a rough estimate of the number of ions in a single GEM CH is 3.5 $\times 10^6$, that of a double GEM is 2.25 $\times 10^6$ and of a triple GEM is 2 $\times 10^6$.
It is very likely that this simple counting of number of ions may not be the only factor determining the transition to streamer from avalanche, but this estimate provides a rough guideline towards the possibility of estimating onset of streamer mode.

\begin{figure}[htbp]
\centering 

\begin{subfigure}{0.5\textwidth}
\centering
\includegraphics[width=.8\linewidth]{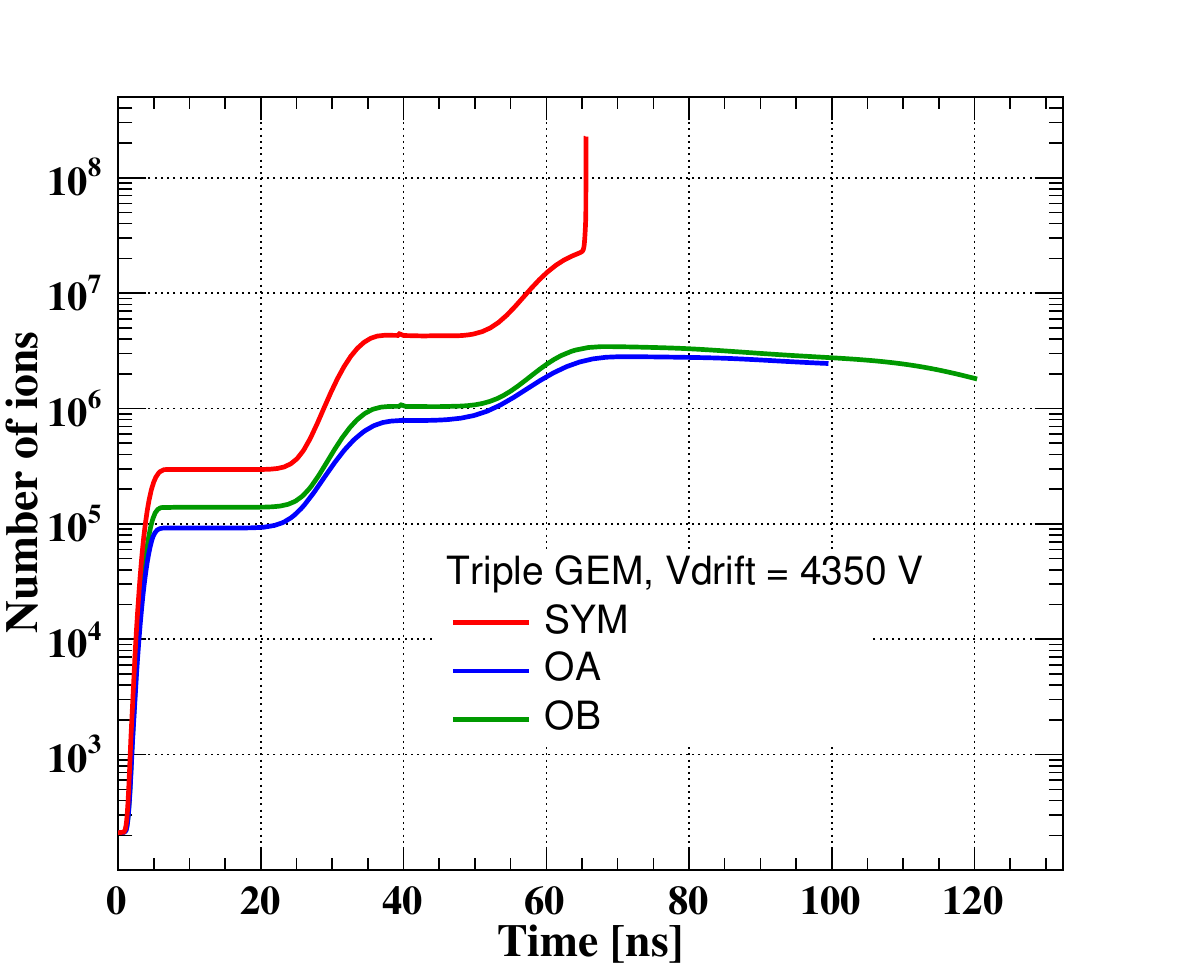}
\caption{}
\end{subfigure}%
\begin{subfigure}{0.5\textwidth}
\centering
\includegraphics[width=.8\linewidth]{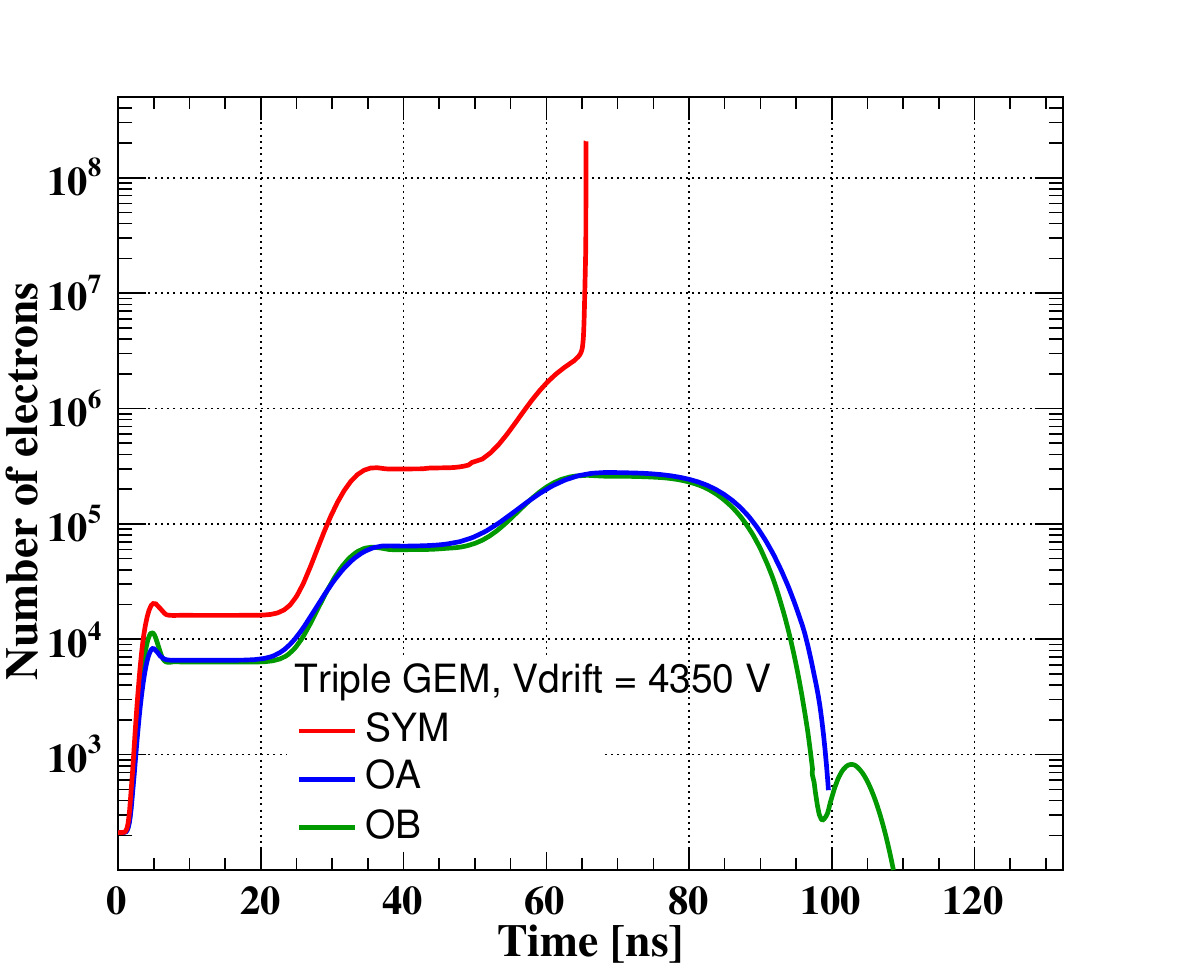}
\caption{}
\end{subfigure}

\caption{\label{TransitionInSYM} Simulated transition from avalanche to streamer modes for SYM configuration in TG1: (a) evolution of number of ions and (b) evolution of number of electrons.}
\end{figure}

In figure \ref{TransitionInSYM}, temporal evolution of electron and ion numbers for TG1 having SYM, OA and OB configurations have been presented.
It may be observed that the lines have a staircase like appearance.
Each step indicates a sharp increase in the number of both ions and electrons due to the passage through a GEM layer.
In the transfer regions, the number of electrons and ions remain unchanged.
In avalanche mode operation, the number of electrons for OA and OB configurations reduces quickly to zero (figure \ref{TransitionInSYM}) since they are collected by the anode at the end of their journey.
Within the time window considered for the simulations, the number of ions for OA and OB start decreasing but is far from vanishing.
The red line of the same figure representing the SYM configuration shows an entirely different behaviour.
The electron numbers increases very rapidly after it reaches $3 \times 10^6$ while the ion numbers goes for a similar rapid increase after reaching a value of $2 \times 10^7$.
This implies a transition to  streamer mode for the SYM configuration for the applied drift voltage of 4350V.
This is expected since electron multiplication in this configuration has been found to be larger than OA and OB.
Despite its less transmission efficiency in comparison to OA, the gain remains high and accumulation of ions happens at a faster rate in SYM.
The charge sharing among the central hole and the hexagons delay the onset of streamer to a certain extent, but as the number of ions in the central hole reaches around 2 $\times 10^6$ (around 10\% of the total number of ions), the transition happens.
It should also be noted from this figure that at the same drift voltage of 4350V, number of ions in OB remains higher than that in OA since the first GEM layer is crossed till the end of the avalanche.
The number of electrons, however, remain almost the same for both configurations.
This picture is consistent with what was noted earlier - the field configuration in OB is likely to lead to more ionizations, but the transmission efficiency of OB is appreciably smaller than that of OA.
As a result, despite the fact that ionization in OA is smaller than that of OB, the gain values are similar.

\begin{figure}[htbp]
\centering 

\begin{subfigure}{0.5\textwidth}
\centering
\includegraphics[width=.8\linewidth]{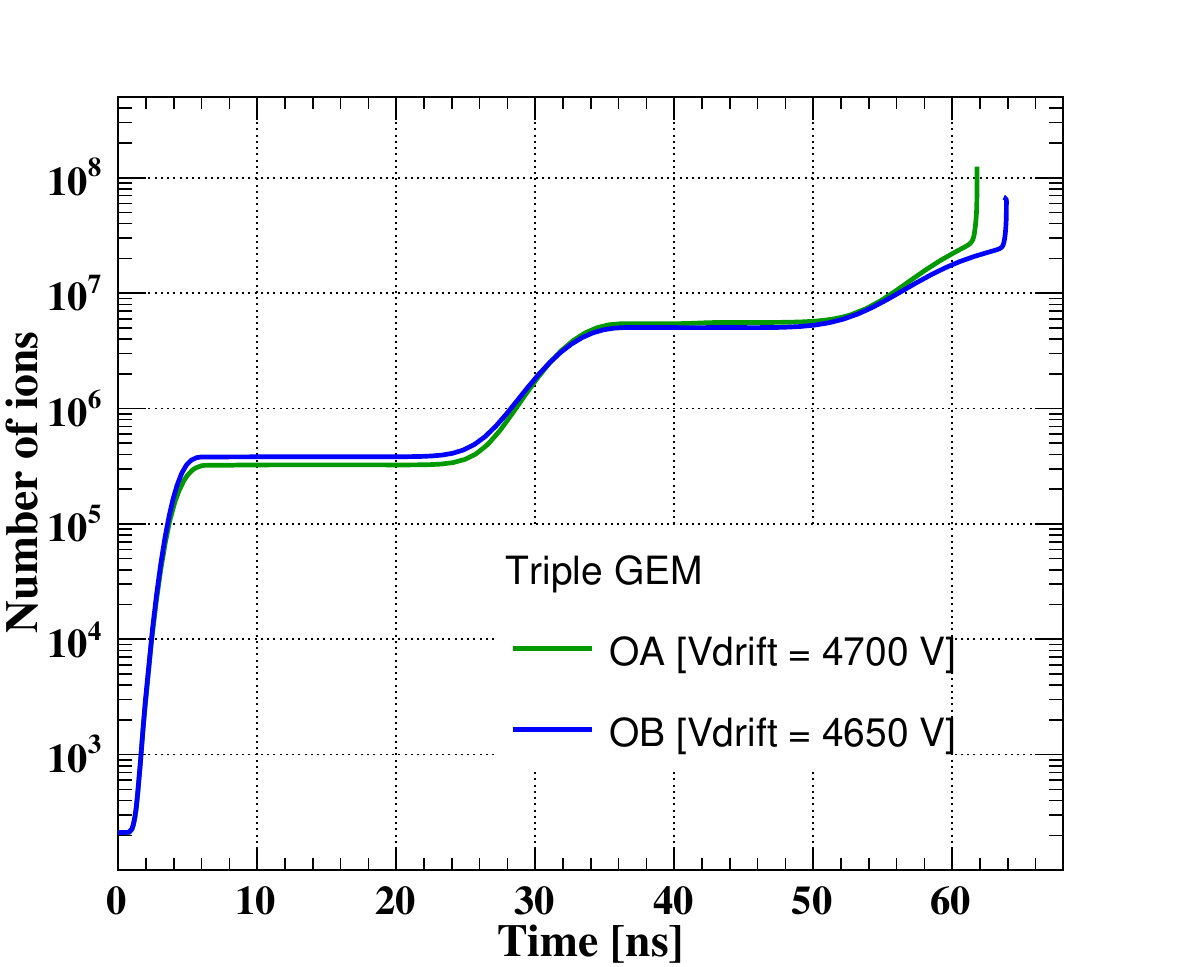}
\caption{}
\end{subfigure}%
\begin{subfigure}{0.5\textwidth}
\centering
\includegraphics[width=.8\linewidth]{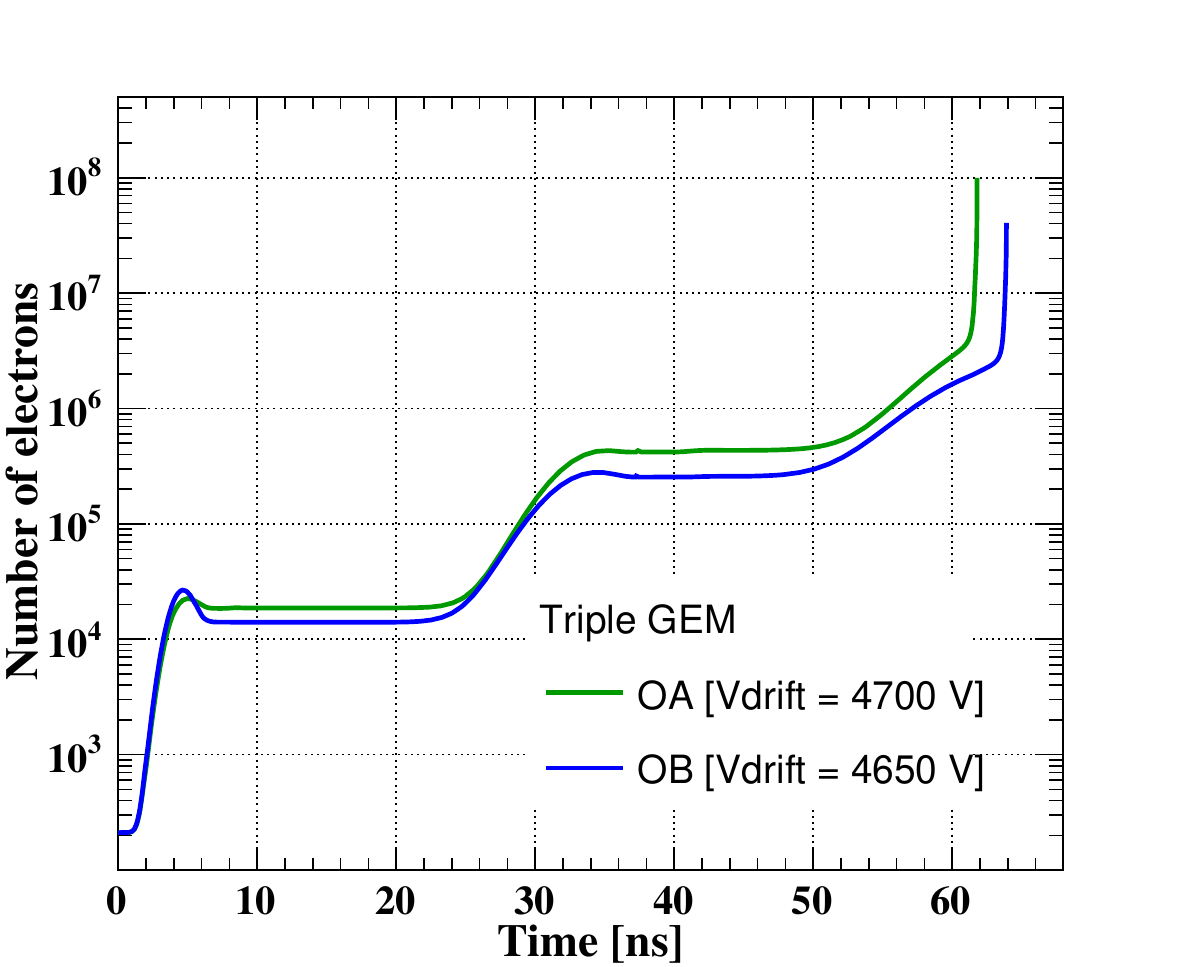}
\caption{}
\end{subfigure}

\caption{\label{TransitionInOAOB} Simulated transition from avalanche to streamer modes for OA and OB configurations in TG1: (a) evolution of number of ions and (b) evolution of number of electrons.}
\end{figure}

The transition to streamer for OA and OB configurations were pursued at higher voltages.
As shown in figure \ref{TransitionInOAOB}, transition in the OA configuration occurs at around 4700V drift voltage, while that for the OB configuration happens slightly earlier, at the drift voltage value of 4650V.
The number of ions for OA at 4700V and OB at 4650V are again close to $2.5 \times 10^7$, a value similar to that observed during the transition to streamer in SYM.
This happens because, due to higher amount of ionization in OB, the number of ions generated in this configuration is larger than that of OA for a given value of drift voltage.
Since transition to streamers is determined by the number of ions in a particular GEM hole, the streamer onset for OB is at a smaller value of drift voltage in comparison to OA.

It should be noted that the drift voltage values estimated to lead to streamers are not expected to match with experimental measurements.
However, it is likely that experimentally measured values will follow the trend of the numerical estimates.

\section{Concluding remarks}
\label{ConRem}
In summary, it may be mentioned that each configuration, SYM, OA and OB, has its own advantages and disadvantages.
SYM has the largest gain but has to be more cautiously dealt with, since it is more prone to discharges.
OA has the largest collection and transmission efficiencies but less multiplication within its holes.
This leads to less gain than that in SYM but the configuration is also more efficient, easier to control and less prone to discharges.
OB has good extraction efficiency, and gains similar to that found in OA.
Like OA, it is also easy to control and less prone to discharges.
From the hydrodynamic simulations, it is observed that both number of ions created and charge sharing play significant roles in deciding the transition from avalanche to streamers.
As a result, accumulation of the similar number of ions in a GEM hole (2-3 $\times 10^6$)triggers the streamer mode for single, double and triple GEM detectors.
The significant enhancement of charge sharing due to additional GEM layers helps in achieving larger gain values before streamers occur.
From this study, it is also found that, between OA and OB, the former is more discharge-safe.
However, careful experiments with very good control of environmental parameters is needed to conclude in this matter.
In order to provide clear guidelines related to the choice of hole geometry, we hope to carry out detailed experiments in the near future and further improve the numerical models presented in this work.

Both the numerical approaches, particle and fluid, have been useful for the present investigations. 
The particle model is capable, self-sufficient and detailed.
It is straightforward to extract any required information easily.
However, since it follows the evolution of each charged particle separately, it is slow.
Inclusion of Coulombic interaction effects leading to space charge is computationally expensive.
As a result, it is difficult to pursue this model when the number of charged particles is large.
On the other hand, the fluid model is most accurate in the region where the particle model faces problems.
In general, the fluid approach lacks in detail and is not self-sufficient since it needs the initial conditions to be supplied from an initial particle description.
In particular, the present axisymmetric model needs the charge sharing information from the particle model to compute the scale factors related to gain estimation.
But, when there are large number of charged particles, the approximation of continuum works fine and inclusion of space charge effect is automatic.
As reflected in the gain plots, the axisymmetric hydrodynamic model adopted here estimates less gain in comparison to the particle model.
Use of three-dimensional fluid model may improve the estimations but will lose the advantage of being much faster than the particle model.

\section{Acknowledgments}
The authors would like to acknowledge Dr. Fabio Sauli for his encouragement in carrying out a study related to charge sharing in GEM-based detectors.
The infrastructural support received from the Saha Institute of Nuclear Physics and the Adamas University is also being acknowledged by the authors.
This work has partly been performed in the framework of the RD51 Collaboration.
We wish to acknowledge the members of the RD51 Collaboration for their help and suggestions.


\end{document}